\documentclass[accepted]{uai2024}

\usepackage[american]{babel}

\usepackage{natbib} 
\usepackage{mathtools} 
\usepackage{booktabs} 
\usepackage{tikz} 

\usepackage{wrapfig}
\usepackage{hyperref}

\newtheorem{definition}{Definition}
\newtheorem{lemma}{Lemma}
\newtheorem{example}{Example}
\usepackage{amsmath}
\usepackage{amsfonts}
\usepackage{subcaption}
\usepackage{graphicx}
\usepackage{algorithm}
\usepackage{algorithmic}
\usepackage{soul} 
\usepackage{url} 

\ifodd 1

\newcommand{\com}[1]{\textbf{\color{red}(COM: #1)}} 
\else

\newcommand{\com}[1]{}
\fi

\newtheorem{theorem}{Theorem}
\newtheorem{proof}{Proof}

\title{Privacy-Aware Randomized Quantization via Linear Programming}

\author[1]{Zhongteng~Cai}
\author[1]{Xueru~Zhang}
\author[1,2]{Mohammad~Mahdi~Khalili}
\affil[1]{%
    Department of Computer Science and Engineering\\
    Ohio State University\\
    Columbus, Ohio, USA
}
\affil[2]{
    Yahoo Research\\
    New York, New York, USA
}

\begin{document}
\maketitle 

\begin{abstract}
Differential privacy mechanisms such as the Gaussian or Laplace mechanism have been widely used in data analytics for preserving individual privacy. However, they are mostly designed for continuous outputs and are unsuitable for scenarios where discrete values are necessary. Although various quantization mechanisms were proposed recently to generate discrete outputs under differential privacy, the outcomes are either biased or have an inferior accuracy-privacy trade-off. In this paper, we propose a family of quantization mechanisms that is unbiased and differentially private. It has a high degree of freedom and we show that some existing mechanisms can be considered as special cases of ours. To find the optimal mechanism, we formulate a linear optimization that can be solved efficiently using linear programming tools. Experiments show that our proposed mechanism can attain a better privacy-accuracy trade-off compared to baselines.



\end{abstract}

\section{Introduction}
\label{sec:intro}

Differential privacy (DP) has become a de facto standard for preserving individual data privacy in data analysis, ranging from simple tasks such as data collection and statistical analysis to complex machine learning tasks \citep{wang2020differentially,jayaraman2019evaluating,zhang2018improving,zhang2018recycled,zhang2019recycled,zhang2022differentially,liu2021robust,khalili2021designing,khalili2021improving,hopkins2022efficient}. It centers around the idea that the output of a certain mechanism or computational procedure should be statistically similar given singular changes to the input, thereby preventing meaningful inference from observing the output. Although many DP mechanisms such as Gaussian mechanism \citep{gaussian}, Laplace mechanism \citep{laplace}, etc., have been proposed to date to preserve individual privacy for different computational tasks, they are mostly designed for continuous outputs over the real numbers and are unsuitable for scenarios where discrete outputs are necessary.  

Indeed, keeping outputs discrete is desirable and even necessary for many applications. For example, representing real numbers on a finite computer requires data discretization, but naively using finite-precision rounding may compromise privacy~\citep{least_bit}. Real-valued outputs can induce high communication overheads, and compressing the continuous inputs to discrete and bounded outputs may be necessary for settings with bandwidth bottlenecks, e.g., federated learning \citep{fedpaq,jin2024performative}. Moreover, continuous outputs are incompatible with cryptographic primitives such as secure aggregation~\citep{secagg}. It is thus essential to develop DP mechanisms that generate discrete outputs while preserving privacy.  

To tackle the challenges mentioned above, many discrete DP mechanisms have been proposed, e.g.,~\citep{discrete-gaussian, skellam, dis_gau_fed}. However, the outputs generated by these mechanisms may be \textit{biased} under truncation. Because in many applications such as machine learning, survey data collection, etc., it is often crucial to maintain the \textit{unbiasedness} of private outputs, these approaches may not be desirable. For instance, when differentially private gradients are used to update machine learning models, keeping them unbiased helps the model gets updated towards the optimal solution and converges faster~\citep{optimization}.


To the best of our knowledge, only a few works proposed mechanisms that can generate discrete unbiased outputs under DP. This includes 1) \textit{Minimum Variance Mechanism} (\textsf{MVU})~\citep{mvu}, which samples outputs from discrete alphabets and achieves the optimal utility by optimizing both the sampling probabilities and output alphabets. However, as the size of output alphabet increases, solving this optimization problem can be particularly challenging and the unbiasedness constraint must be relaxed; 2) \textit{Randomized Quantization Mechanism} (\textsf{RQM})~\citep{rqm} which randomly maps inputs to closest pair of sampled bins. However, \textsf{RQM} assumes uniformly distributed bins and has only three hyperparameters that can be tuned, hence has smaller search space for hyperparameters to achieve good privacy-accuracy trade-off compared with \textsf{MVU}; 3) \textit{Poisson Binomial Mechanism} (\textsf{PBM})~\citep{pbm} which generates unbiased estimators by mapping inputs to a discrete distribution with bounded support. However, \textsf{PBM} has inferior flexibility and utility-privacy trade-off than \textsf{RQM} because it has fewer hyperparameters; 4) other DP mechanisms such as \textit{Distributed Discrete Gaussian Mechanism}~\citep{dis_gau_fed} and \textit{Skellam Mechanism}~\citep{skellam} are unbiased on the unbounded support. However, they have to be truncated when combined with secure aggregation protocols, which will produce biased outputs.


This paper proposes a novel randomized quantization mechanism with discrete, unbiased outputs under DP guarantee. Importantly, our mechanism ensures unbiasedness regardless of the number of output bits; it is a general framework and the existing mechanism \textsf{RQM} can indeed be considered as a special case of ours. Specifically, given a set of quantization bins $B_1 <B_2< \cdots< B_m$, discrete DP mechanism maps the continuous input $x$ to one of these bins. Our mechanism first samples two bins from the left and the right side of the input based on a pre-defined \textit{selection distribution}, and then outputs one of the bins with unbiased expectation. For an example where $m=4$ and $x \in [B_2, B_3)$. Our mechanism first randomly selects one bin on the left of $x$ (e.g., $B_1$) and another bin on the right (e.g., $B_3$) according to a selection distribution, then randomly outputs either $B_1$ or $B_3$ while preserving unbiasedness.
The key is to carefully design selection distributions that maximize the accuracy of quantized outputs subject to DP constraint. Although this problem can be easily formulated as a non-linear constraint optimization, we propose a method that turns such non-linear optimization into a linear program that can be solved efficiently using linear programming tools. Experiments on both synthetic and real data validate the effectiveness of the proposed method. The code repository for this work can be found in \url{https://github.com/osu-srml/DP_Linprog}.


Our contribution can be summarized as follows:
\begin{enumerate}[leftmargin=*]
    \item We propose a family of differentially private quantization mechanisms that generate discrete and unbiased outputs.
    \item We theoretically quantify the privacy and accuracy of the exponential randomized mechanism (\textsf{ERM}), a special case of our proposed mechanism where selection distribution is based on DP exponential mechanism. 
    \item We design a linear program to find the optimal selection distribution of our mechanism, resulting in the optimal randomized quantization mechanism
(\textsf{OPTM}), which attains a better accuracy-privacy trade-off.
    \item We conduct experiments on various tasks to show our mechanisms, including both \textsf{ERM} and \textsf{OPTM}, attain superior performance than baselines.
\end{enumerate}

\section{Related Works}
\label{sec:related_work}

\paragraph{Discrete differential privacy.} Various discrete DP mechanisms have been proposed for discrete inputs to make them differentially private. For example, both \textit{Discrete Laplace Mechanism}~\citep{discrete-laplace} and \textit{Discrete Gaussian Mechanism}~\citep{discrete-gaussian} add noises to the inputs sampled from discrete distributions, which are commonly used for tasks when with discrete inputs~\citep{abowd}. The \textit{Snapping Mechanism}~\citep{least_bit} truncates and rounds the inputs and Laplace noises based on floating-point arithmetic, but it inevitably diminishes accuracy~\citep{discrete-gaussian}. \textit{Communication-limited Local Differential Privacy} (CLDP) mechanism~\citep{shuffled} works with a trusted shuffler in federated learning to generate compressed and private updates from clients. However, it cannot be tuned to adopt different communication budgets. \textit{Skellam Mechanism}~\citep{skellam} add noises sampled from Skellam distribution to achieve performance comparable with the continuous Gaussian mechanism, but is subject to biased output when combined with privacy-protection protocols in federated learning such as secure aggregation. In contrast, \textit{Poisson Binomial Mechanism}~\citep{pbm} encodes the inputs inside the Binomial distribution to generate unbiased outputs, and it can achieve better privacy while decreasing communication costs, and is also compatible with secure aggregation.

\paragraph{Private quantization.} 
Previous works have utilized data compression methods such as quantization to compress the data in applications with communication or bandwidth bottlenecks. One example is federated learning where a central server needs to repeatedly collect local model updates from distributed clients for training the global model~\citep{fedpaq, dadaquant}. Another example is large language models where the computation overheads may be reduced by compressing the model parameters~\citep{compress-model, post-train-quant}.
By mapping the continuous inputs to the closest discrete outputs within a finite set, the quantization process can effectively represent the data with reduced communication overhead. 

While methods were proposed in prior works to quantize data under a certain privacy constraint, they often treat privacy and quantization separately~\citep{vqsgd, dis_gau_fed}, i.e., privatizing the data first and then quantize the private data. 
Recent works attempt to design discrete DP mechanisms leveraging quantization to simultaneously compress data and protect privacy. For instance, \cite{mvu} proposed \textit{Minimum Variance Mechanism} (\textsf{MVU}), which first quantizes inputs with discrete bins and then maps the unbiased quantization results to output alphabets according to a probability matrix. \textsf{MVU} optimizes the probability matrix to minimize accuracy loss while preserving privacy. \textsf{I-MVU}~\citep{i-mvu} extends \textsf{MVU} by 
designing a new interpolation procedure to attain better privacy for high-dimensional vectors. \cite{rqm} proposed \textit{Randomized Quantization Mechanism} (\textsf{RQM}), which subsamples from uniformly distributed bins and performs randomized quantization to output an unbiased result. 

Compared to prior works, we propose a more general family of quantization DP mechanisms that enables non-uniform quantization. It has a high degree of freedom and the optimal mechanism can be found efficiently by linear programming tools. We also show theoretically and empirically that our mechanism can attain a better privacy-accuracy trade-off. 



\section{Problem Formulation}
\label{sec:mechanism_design}
Consider a quantization mechanism $$\mathcal{M}: \mathcal{X}\rightarrow \{B_1, B_2, \cdots , B_m\}$$ used for quantizing a scalar $x \in [-c,c]:=\mathcal{X}$, where $$-c-\Delta=B_1< B_2 < \cdots < B_m=c+\Delta,$$ $m$ is the number of quantization bins, $\Delta\geq0$ extends the range of output. Note that $m$ bins here are not necessarily uniformly distributed. Our goal is to design $\mathcal{M}$ (including bin values $B_1,\cdots,B_m$ and $\Delta$) that is 1) differentially private; 2) unbiased, i.e., $\mathbb{E}(\mathcal{M}(x)) = x, \forall x$; and 3) accurate with the mean absolute error $\mathbb{E}(|\mathcal{M}(X)-X|)$ minimized. Let the capital letter $X$ denote the random variable of input and the small letter $x$ the corresponding realization. 

\subsection{Background: differential privacy}


Differential privacy \citep{dwork2006differential}, a widely used notion of privacy,
ensures that no one by observing the computational outcome can infer a particular individual’s data
with high confidence. Formally, we say a randomized algorithm $\mathcal{M}(\cdot)$ satisfies 
$\epsilon$-differential privacy (DP) if for any two datasets $D$ and $D'$ that are different in at most one individual's data and for any set of
possible outputs $S\subseteq \text{Range}(\mathcal{M})$, we have,
$$
	{\Pr}\{\mathcal{M}(D)\in S\} \leq \exp\{\epsilon\}\cdot {\Pr}\{\mathcal{M}(D')\in S\}.
$$
where $\epsilon\in [0,\infty)$ is called privacy loss
and can serve as a proxy for privacy leakage; the smaller $\epsilon$ implies a stronger privacy guarantee. Intuitively, for sufficiently small $\epsilon$, DP implies that the distribution of output remains almost the same if one individual's data changes in the dataset, and an attacker cannot reconstruct input data with high confidence after observing the output of mechanism $\mathcal{M}$.

Many mechanisms have been developed in the literature to satisfy differential privacy. One that is commonly used for scenarios with discrete outputs is \textit{exponential mechanism} \citep{mcsherry2007mechanism}, as defined below.

\begin{definition}[Exponential Mechanism]\label{def:em}
Let the set of all possible outcomes of mechanism $\mathcal{M}$ be $\mathcal{O}= \{o_1, \cdots, o_{\hat{n}}\}$. Let  $v:\mathcal{O}\times \mathcal{D}\rightarrow \mathbb{R}$ be a score function with a higher value of $v(o_i,D)$ indicating output $o_i$ is more desirable under dataset $D$. Let $\delta =  \max_{i,D,D'} |v(o_i,D)-v(o_i,D')|$ be the sensitivity of score function, where $D$ and $D'$ are two datasets differing in at most one individual's data.  
Then, exponential mechanism  $\mathcal{M}:\mathcal{D}\rightarrow \mathcal{O}$ that satisfies  $\epsilon$-differential privacy  selects  $o_i \in \mathcal{O}$  with  probability  
$$
		{\Pr}\{\mathcal{M}(D) = o_i\} = \frac{\exp\left\{\epsilon \cdot \frac{v(o_i,D)}{2\delta}\right\}}{\sum_{j=1}^{\hat{n}}\exp\left\{\epsilon \cdot \frac{v(o_j,D)}{2\delta}\right\}}.
$$
\end{definition}

\subsection{Proposed quantization mechanism}\label{subsec:mechanism}
Next, we present our mechanism $\mathcal{M}$ that quantizes input with DP guarantee.
Given a set of bins $\{B_1,B_2,\cdots, B_m\}$, $\mathcal{M}$ takes the following steps to quantize a scalar $x$:
\begin{enumerate}[leftmargin=*]
    \item For any $x$, select two bins $B_l, B_r\in \{B_1,B_2,\cdots, B_m\}$ randomly based on a pre-defined \textit{selection distribution}, with $B_l \leq x$ located on the left side of $x$ and $B_r > x$ on the right side of $x$. In other words, if $x\in [B_j,B_{j+1})$, then $l \in \{1,\ldots,j\}$ and $r \in \{j+1,\ldots,m\}$.
    \item Then, $\mathcal{M}$ randomly outputs either $B_l$ or $B_r$ according to
\begin{equation}\label{equ:dither}
            \mathcal{M}(x) = 
                \begin{cases}
                    B_l, & \text{with probability (w.p.)}~ \frac{B_r-x}{B_r-B_l}; \\
                    B_r, & \text{with probability (w.p.)}~ \frac{x-B_l}{B_r-B_l}.
                \end{cases}
        \end{equation}
\end{enumerate}
Given \eqref{equ:dither}, it is easy to verify that the mechanism $\mathcal{M}$ is unbiased, i.e., $\mathbb{E}(\mathcal{M}(x)) = x, \forall x$.  Our goal is to design \textit{selection distribution} in the first step such that the mean absolute error  $\mathbb{E}(|\mathcal{M}(X)-X|)$ is minimized. In this paper, we assume bin values $\{B_1,\cdots, B_m\}$ are symmetric unless otherwise stated, i.e., $B_{i} = -B_{m+1-i}$, $\forall i\in [m]$.   

\begin{figure}[ht]
\begin{center}
\centerline{\includegraphics[width=0.8\columnwidth]{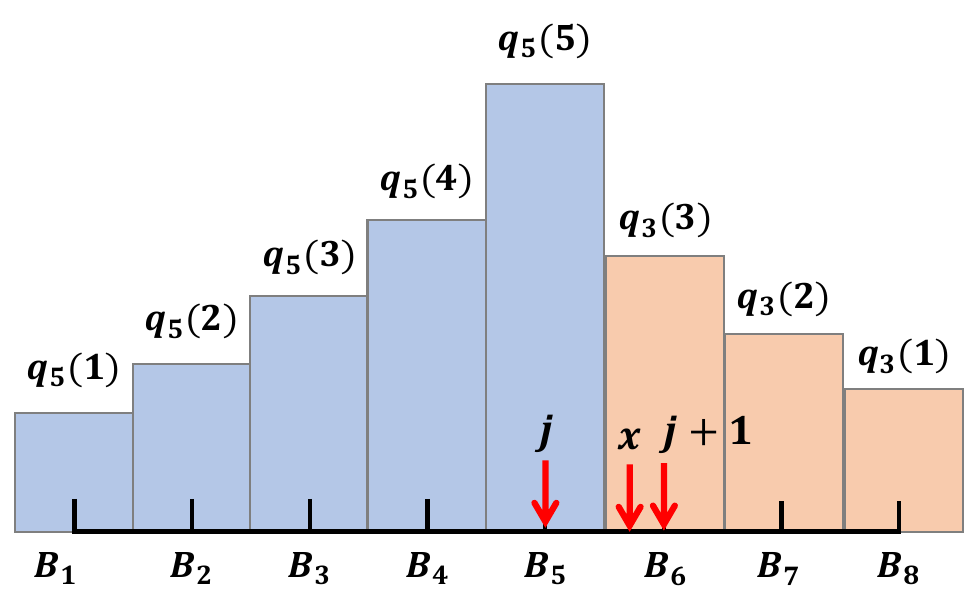}}
\caption{An example of selection distribution}
\label{fig:dis}
\end{center}
\vskip -0.4in
\end{figure}

\paragraph{Selection distribution.} It determines the probability of selecting one bin on the left (or right) of the input $x$ in the first step of our mechanism. Assume $x \in [B_j, B_{j+1})$, then we will select the left index $l\in \{1,\cdots,j\}$ and the right index $r\in \{j+1,\cdots,m\}$. Let $L_j$ and $R_j$ be the random variables associated with the left index $l$ and right index $r$, respectively, when input $x \in [B_j, B_{j+1})$. Since the probability mass functions (PMF) of both $L_j$ and $R_j$ depend on the value of $j$, we use the following two functions $q_j,q_{m-j}$ to denote their PMF:
\begin{align*}
  &\Pr\{L_j=i\}:=q_j(i),&i\in \{1,\cdots,j\}\\
  &\Pr\{R_j = i\}:=q_{m-j}(m+1 - i),&i \in \{j+1,\ldots,m\}
\end{align*}
Note that $q_1(1) = 1$. See Figure \ref{fig:dis} for the illustration.


Since both $q_j(\cdot), j\in \{1,2,\cdots,m\}$ and $\{B_1,\cdots, B_m\}$ are the parameters of mechanism $\mathcal{M}$, we need to design them carefully to minimize the absolute error while satisfying DP constraint. We introduce details of finding these parameters in Section~\ref{subsec:gen}. Given  $q_j(\cdot)$ and $\{B_1,\cdots, B_m\}$, Algorithm~\ref{alg:gen} summarizes our mechanism $\mathcal{M}$.


\begin{algorithm}[tb]
\caption{Proposed quantization mechanism $\mathcal{M}$}
   \label{alg:gen}
\begin{algorithmic}[1]
   \STATE {\bfseries Input:} bin values ${B_1, \cdots, B_m}$, input $x \in [B_j, B_{j+1})$, PMF $q_j,q_{m-j}$ of $L_j$ and $R_j$. 
   \STATE $l \gets i$ w.p. $q_j(i)$, $i \in \{1,\cdots,j\}$.
   \STATE $r \gets i$ w.p. $q_{m-j}(m+1-i)$, $i \in \{j+1,\cdots,m\}$.
   \STATE $\mathcal{M}(x) \gets B_l$ w.p. $\frac{B_r-x}{B_r-B_l}$, and $B_r$ w.p. $\frac{x-B_l}{B_r-B_l}$.
   \STATE {\bfseries Output} $\mathcal{M}(x)$
\end{algorithmic}
\end{algorithm}


\subsection{Special cases}\label{subsec:special}
Section~\ref{subsec:mechanism} presents a general framework for quantization. Indeed, some existing mechanisms proposed in prior works can be regarded as a special case of ours, as detailed below.

\textbf{Randomized Quantization Mechanism (\textsf{RQM})  .}~~
It is proposed by \citet{rqm} and is a special case of ours. Specifically, $\forall x \in [-c, c]$, \textsf{RQM} randomly outputs one bin from $\{B_1,\cdots,B_m\}$ with $$B_i = -\Delta-c + (i-1)\frac{2c+2\Delta}{m-1}, ~ i\in [m], $$ That is, interval $[-c-\Delta, c+\Delta]$ is divided uniformly into $m$ bins. This differs from ours where we enable non-uniformly distributed bins and the bin values are parameters to be optimized (see details in Section~\ref{subsec:gen}). 

To quantize $x$, \textsf{RQM} first selects a subset of bins: $B_1$ and $B_m$ are selected with probability $1$, while among the rest $m-2$ bins $\{B_2,\cdots, B_{m-1}\}$, each of them is selected independently with probability $q<1$. Given the selected bins, the one closest to $x$ on the left (resp. right) side is denoted as $B_l$ (resp. $B_r$). Finally, \textsf{RQM} selects either $B_l$ or $B_r$ as the output randomly based on Eq.~\eqref{equ:dither}. It turns out that \textsf{RQM} is a special case of our mechanism where selection distribution follows a \textit{Geometric distribution} with parameter $q$, i.e.,  
\begin{equation*}
    q_j(i) = 
\begin{cases}
    (1-q)^{j-1},& \text{if $i = 1$} \\
    q{(1-q)}^{j-i}, & \text{if $1 < i \leq j$}.
\end{cases}
\end{equation*}




\paragraph{Exponential Randomized Mechanism (\textsf{ERM}).}
Inspired by the classic \emph{Exponential Mechanism} (Definition~\ref{def:em}), we can propose \textsf{ERM} which outperforms \textsf{RQM} (see the comparison in Section~\ref{sec:res}) but can still be regarded as a special case of our proposed mechanism. Under \textsf{ERM}, bins are symmetric and satisfy $B_i = -B_{m+1-i}, \forall i \in [m]$.  \textsf{ERM} uses a distribution similar to the exponential mechanism for the selection distribution. Specifically, for input $x\in [B_j,B_{j+1})$, PMF $\Pr\{L_j  = i\} = q_j(i)$ in \textsf{ERM} depends on the distance between bin $B_i$ and $B_j$ and 
$\Pr\{R_j  = i\} = q_{m-j}(m+1-i)$ depends on the distance between bin $B_i$ and $B_{j+1}$. In other words, \textsf{ERM} uses the following selection distribution: 
\begin{equation}\label{equ:exp}
    q_j(i) = 
        \frac{\exp\left\{\frac{\gamma(B_i-B_j)}{2(B_j-B_1)}\right\}}{\sum_{k=1}^{j}\exp\left\{\frac{\gamma(B_k-B_j)}{2(B_j-B_1)}\right\}},
\end{equation}
where $\gamma$ is a hyperparameter impacting both the privacy and accuracy of $\mathcal{M}$. After obtaining the realizations of $L_j$ and $R_j$, \textsf{ERM} uses Eq.~\eqref{equ:dither}to determine the final output. 

Next, we provide privacy and accuracy analysis for \textsf{ERM}. Theorem~\ref{thm:erm_privacy} below provides an upper bound for privacy loss. 

\begin{theorem}[Privacy loss of \textsf{ERM}]\label{thm:erm_privacy}
Assume the interval $[-c-\Delta, c+\Delta]$ is divided uniformly into $m$ bins, i.e., $$B_i = -\Delta-c + (i-1)\frac{2c+2\Delta}{m-1}, ~ i\in [m].$$ 
Then \textsf{ERM} satisfies DP with privacy loss 
\begin{equation}\label{eq:privacy_erm}
 \epsilon < \gamma + \log \frac{2m(c+\Delta)}{c}.   
\end{equation}
\end{theorem}

The upper bound~\eqref{eq:privacy_erm} implies that the privacy loss is an increasing function in the number of bins $m$ and parameter $\gamma$. It is worth noting that according to \citep{rqm}, the privacy loss of \textsf{RQM} is bounded by $$\log\left(\frac{2(1-q)^2(c+\Delta)}{\Delta}\right) +m\log\frac{1}{1-q}.$$ This shows that the privacy loss under \textsf{RQM} also increases in $m$ at the rate of $\mathcal{O}(m)$. In contrast, our \textsf{ERM} has a better privacy loss that increases in $m$ at the rate of $\mathcal{O}(\log m)$.

The next theorem provides an upper bound for the expected absolute error of \textsf{ERM}. 
\begin{theorem}[Error of \textsf{ERM}]\label{thm:erm_error}
Under the same bins as Theorem~\ref{thm:erm_privacy}, 
the expected absolute error of \textsf{ERM} is bounded:
\begin{equation*}
 \mathbb{E}\left(|\mathcal{M}(x)-x|\right) \leq \frac{4}{\gamma}\log\left(m\right)\left(c+\Delta\right)+\frac{2c+2\Delta}{m-1}. 
\end{equation*}
\end{theorem}
The bound implies that when the extended range $\Delta$ increases or the privacy budget parameter $\gamma$ decreases (stricter privacy protection), the performance loss will also increase.


\section{Optimal Mechanism}\label{subsec:gen}

Section~\ref{subsec:mechanism}  introduced the general framework of our quantization mechanism. With different bin values $\{B_1,\cdots,B_m\}$ and selection distributions $q_j,q_{m-j}$, we will end up with different mechanisms and we discussed two special cases in Section~\ref{subsec:special}. In this section, we explore how to find the optimal mechanism by tuning these parameters. We call the quantization mechanism under the optimal parameter configuration ``{\textbf{OPT}imal randomized quantization \textbf{M}echanism (\textsf{OPTM})}." Before introducing \textsf{OPTM}, we first quantify privacy loss and mean absolute error of our mechanism under a given bin values $\{B_1,\cdots, B_m\}$ and selection distributions.

\subsection{Performance measure}\label{subsec:measure}

 Given bin values $\{B_1,\cdots,B_m\}$ and selection distributions $q_j,q_{m-j}$, we can find the \textit{output distribution} $\Pr\{\mathcal{M}(x) = i\}, i \in [m]$ for any input $x$. Let $p(x, i):= \Pr\{\mathcal{M}(x) = i\}$ be the probability that the output of the mechanism $\mathcal{M}$ for an input $x$ is $B_i$. Then, the probability that $\mathcal{M}$ outputs bin $B_l$ on the left of $x$ can be calculated by the law of total probability as follows,
\begin{equation*}\label{equ:p_left}
\resizebox{0.43\textwidth}{!}{$\displaystyle   p(x, l)=\Pr\{L_j=l\} \sum_{m \geq r\geq j+1}\left( \Pr\{R_j = r \} \frac{B_r-x}{B_r-B_l} \right).$}  
\end{equation*}
Similarly, for a bin $B_r$ on the right side of $x$, we have
\begin{equation*}\label{equ:p_right}
\resizebox{0.43\textwidth}{!}{$  \displaystyle  p(x, r)=\Pr\{R_j = r\} \sum_{1\leq l \leq j}\left( \Pr\{L_j = l\}\frac{x-B_l}{B_r-B_l} \right). $}
\end{equation*}
Hence, the output probability of each bin $B_i$ is given by:
{
\begin{align}\label{equ:output}
 \resizebox{0.43\textwidth}{!}{$  p(x,i) =\begin{cases}
    \displaystyle        q_j(i)  \sum_{r \in [j+1,m]} \left( q_{m-j}(m-r+1) \frac{B_r-x}{B_r-B_{i}}  \right), & \text{if $B_{i} \leq x$} \\
            \displaystyle q_{m-j}(m+1 - i )  \sum_{l \in [1, j]}\left( q_j(l) \frac{x-B_l}{B_{i}-B_l} \right), & \text{o.w.}
        \end{cases}$}
\end{align}}
\paragraph{Performance measure.} With the output distribution computed above, we can quantify the mean absolute error (MAE) of a mechanism $\mathcal{M}$ as follows,
\begin{equation}\label{eq:objective}\mathbb{E}\left(\lvert\mathcal{M}(X)-X\rvert\right) = \mathbb{E}_X\left(\sum_{i \in [m]} p(X,i) \lvert B_{i}-X\rvert\right).
\end{equation} 
To satisfy differential privacy, the output distribution with bounded privacy loss $\epsilon$ should satisfy: 
\begin{align}\label{equ:con}
    \frac{p(x,i)}{p(x^{\prime},i)} \leq e^{\epsilon}, \quad \forall x,x' \in [-c, c], i \in [m].
\end{align}
Our goal is to design parameters of $\mathcal{M}$, including $\Delta$, bin values $\{B_{1},\cdots,B_m\}$ and especially selection distributions $q_j, q_{m-j}$, such that MAE is minimized subject to bounded privacy loss $\epsilon$. Note that since $m$ determines the number of bits for quantizing $x$ (e.g., 2 bits equals $m=4$), we assume $m$ is pre-defined and is not a variable to be optimized.

\subsection{\textsf{OPTM} as a linear program}

The problem of finding the optimal parameters of $\mathcal{M}$ can be formulated as an optimization. Our goal is to simplify the optimization as a \textit{linear program} that can be efficiently solved using linear programming tools. Next, we first derive a linear upper bound of the objective function \eqref{eq:objective}. Then, we describe how to turn DP constraint~\eqref{equ:con} into linear constraints. Finally, we show how to reduce the complexity when the number of output bits is large with another set of constraints.

\paragraph{Linear upper bound for MAE.} Eq.~\eqref{eq:objective} shows that the mean absolute error is a non-linear function of $q_j(i)$. However, we can find a linear upper bound of it and use it as a proxy, as detailed below. 
\begin{lemma}\label{lemma:x_error}
    For any input $x \in [B_{j}, B_{j+1})$, we have, 
\begin{equation*} \mathbb{E}\left(|\mathcal{M}(x)-x|\right) \leq \frac{1}{2}\Big(\zeta_{m-j} + \left(B_{j+1}-B_{j}\right) + \zeta_{j}\Big), 
    \end{equation*}
    where $\zeta_n = \sum_{i \in [n]} q_n(i) (B_{n}-B_{i})$. 
\end{lemma}
If we know the distribution of $X$, we can use Lemma~\ref{lemma:x_error} to further find a linear upper bound of  $\mathbb{E}(|\mathcal{M}(X)-X|)$. An example for uniformly distributed $X$ is given in Theorem~\ref{thm:uniform}.


\begin{theorem}\label{thm:uniform}
Suppose input $x\in[-c,c]$ follows uniform distribution, $\mathbb{E}(|\mathcal{M}(X)-X|)$ can be upper bounded by
\begin{equation}
\label{equ:opti}
\resizebox{0.43\textwidth}{!}{$\displaystyle
 \min_{q_j(i)} \sum_{s\leq n\leq  t+1} \big(\min(c, B_{n})-\max(-c, B_{n-1})\big) \big(\zeta_{n-1} + \zeta_{m-n+1}\big),$}  
\end{equation}
where $\zeta_n = \sum_{i \in [n]} q_n(i) (B_{n}-B_{i})$. $B_{s-1}\in [-c-\Delta,-c)$ and $B_{t+1}\in (c, c+\Delta]$ are two bins fall in extended range, $B_s<B_t$ are bins in $[-c,c]$ closest to $-c$ and $c$, respectively.
\end{theorem}


For more general cases with partially known, non-uniformly, and even asymmetric distributed input $X$, our mechanism can still be adapted. 

Specifically, we first change the original definition of selection distribution in Section~\ref{subsec:mechanism} to the following:
\begin{align*}
  &\Pr\{L_j = i\}:=q_j^{(l)}(i),&i\in \{1,\cdots,j\}\\
  &\Pr\{R_j = i\}:=q_{m-j}^{(r)}(m+1-i),&i \in \{j+1,\ldots,m\}
\end{align*}
Both $q_j^{(l)}(\cdot), q_{m-j}^{(r)}(\cdot)$ for all possible $j \in [m]$ are parameters that need to be tuned. Then we can derive a linear upper bound of the mean absolute error (MAE) by extending Lemma~\ref{lemma:x_error} and Theorem~\ref{thm:uniform}. Theorem~\ref{thm:extend} below shows the result for non-uniformly distributed $X$ and asymmetric bins.

\begin{theorem}\label{thm:extend}
Suppose input $x\in[-c,c]$ follows any distribution, $\mathbb{E}(|\mathcal{M}(X)-X|)$ can be upper bounded by
\begin{equation}
\label{equ:opti}
\resizebox{0.43\textwidth}{!}{$\displaystyle
 \min_{q_j^{(l)}(i), q_{m-j}^{(r)}(i)} \sum_{i=s-1}^{t}  (\zeta_{m-i}^{(r)}+B_{i+1}-B_{i}+\zeta_{i}^{(l)}) \int_{\max(B_i, -c)}^{\min(B_{i+1}, c)} f_X(x) dx,$}  
\end{equation}
where $\zeta_{m-j}^{(r)} = \sum_{i \in \{j+1, \dots, m\}} q_{m-j}^{(r)}(m-i+1) (B_{i}-B_{j+1})$, $\zeta_j^{(l)} = \sum_{i \in [j]} q_j^{(l)}(i) (B_{j}-B_{i})$. $B_{s-1}\in [-c-\Delta,-c)$ and $B_{t+1}\in (c, c+\Delta]$ are two bins fall in extended range, $B_s<B_t$ are bins in $[-c,c]$ closest to $-c$ and $c$, respectively. $f_X(x)$ is the probability density function of $X$.
\end{theorem}
Note that the upper bound in Theorem~\ref{thm:extend} only depends on density $f_X(x)$ through the integral $\Pr(B_i \leq X < B_{i+1}) = \int_{B_i}^{B_{i+1}} f_X(x)dx$, which is easier to know (compared to density itself) and can be estimated from samples.

\paragraph{Linear differential privacy constraint.}  To satisfy $\epsilon$-DP, constraint~\eqref{equ:con} can be equivalently written as
\begin{equation}\label{equ:privacy_constraint}
\frac{\max_x p(x,i)}{\min_{x^{\prime}} p(x^{\prime},i)} \leq e^{\epsilon}, \quad \forall i     
\end{equation}
However, constraint~\eqref{equ:privacy_constraint} is non-linear and we need to convert it to a linear constraint. To this end, we will first show in Lemma~\ref{lemma:pr_set} that for each $i \in [m]$ and $x \in [-c, c]$, both $\max_x p(x,i)$ and $\min_x p(x,i)$ can be found in a finite set. Such property will then be leveraged to turn constraint \eqref{equ:privacy_constraint} into a set of linear constraints. 
\begin{lemma}\label{lemma:pr_set}
     Assume that $\forall i,j \in [m], j \geq i$: $q_i(i) \geq q_j(i)$, then for all input $x \in [-c, c]$ and each $i \in [m]$: $$\textstyle \max_x p(x, i) \in \overline{\mathcal{S}}_i ~~~~\text{ and }~~~~\min_x p(x, i) \in \underline{\mathcal{S}}_i,$$ where both $\overline{\mathcal{S}}_i$ and $\underline{\mathcal{S}}_i$ are finite sets defined below.
\begin{align*}
\resizebox{0.42\textwidth}{!}{$
\overline{\mathcal{S}}_i= \begin{cases}
  \{q_i(i), q_{m+1-i}(m+1-i)\}, & \text{ if } B_{i} \in [-c,c] \\
 \{ p(-c, i)\} \cup \{p(B_k, i) | -c\leq B_k \leq c\},& \text{ if } B_i < -c.\\
 \{p(c, i)\} \cup \{p(B_k, i) | -c\leq B_k \leq c\},& \text{ if } B_i >c.
\end{cases}$}
\end{align*}
\begin{align*}
\resizebox{0.48\textwidth}{!}{$ \displaystyle
\underline{\mathcal{S}}_i= \begin{cases}
\displaystyle \left\{p(-c,i), p(c,i)\right\}\cup \Big\{ \lim_{x \to B_k} p(x,i) |-c\leq B_k \leq c\Big\}, & \text{ if } B_{i} \in [-c,c] \\
 \displaystyle \left\{p(c,i)\right\}\cup \Big\{ \lim_{x \to B_k} p(x,i) |-c\leq B_k \leq c\Big\},& \text{ if } B_i < -c.\\
\displaystyle \left\{p(-c,i)\right\} \cup \Big\{ \lim_{x \to B_k} p(x,i) |-c\leq B_k \leq c\Big\},& \text{ if } B_i >c.
\end{cases}$}
\end{align*}
where $\lim_{x \to B_k}p(x,i)$ above is calculated as follows
\begin{align*}
\resizebox{0.48\textwidth}{!}{$ \displaystyle
 \lim_{x \to B_k}p(x,i)= 
 \begin{cases}
\displaystyle q_{k-1}(i)  \sum_{r \in [k+1, m]}\bigg( q_{m-k+1}(m-r+1) \frac{B_r-B_k}{B_r-B_{i}} \bigg),& \text{ if }B_i < B_k.\\
\displaystyle q_{m-k}(m+1-i)  \sum_{l \in [1, k-1]}\bigg( q_k(l) \frac{B_k-B_l}{B_i-B_l} \bigg),& \text{ if }B_i > B_k.
 \end{cases} $}  
\end{align*}


    







\end{lemma}
Note that $p(x,i)$ is discontinuous and $\lim_{x \to B_k}p(x,i)$ may not equal to $p(B_k,i)$. Lemma~\ref{lemma:pr_set} shows that for each $i \in [m]$, there is only a finite number of possible values for both $\max_x p(x,i)$ and $\min_x p(x,i)$. Therefore, if  we can ensure $\overline{s} \leq e^\epsilon \cdot \underline{s}$ holds for any $\overline{s} \in\overline{\mathcal{S}}_i$ and $\underline{s} \in\underline{\mathcal{S}}_i$, then privacy constraint~\eqref{equ:privacy_constraint} is also guaranteed to hold. The monotonicity of the output probability between each pair of bins ensures that we can find a finite set of maximal and minimal probabilities. Example 1 uses specific output distributions of $\textsf{ERM}$ to illustrate this.
\begin{example}
    Figure~\ref{fig:erm_dist} shows two probabilities $p(x,3)$ and $p(x,6)$ of $\textsf{ERM}$ when $m=8$. Note that $\lim_{x \to B_3^{+}} p(x,3)=p(B_3,3)=q_3(3)$ and $\lim_{x \to B_3^{-}} p(x,3)=q_6(6)$. We have $\max_x p(x,i)\in \{q_3(3), q_6(6)\}$. When $x$ increases from $B_3$ to $B_4$, or decreases from $B_3$ to $-c$, $p(x,3)$ decreases. When $x$ increases from $B_4$ to $B_5$, $B_5$ to $B_6$, $B_6$ to $c$, $p(x,3)$ also decreases. Hence, we have $\min_x p(x,3) \in \{p(-c,3)$, $\lim_{x \to B_4} p(x,3)$, $\lim_{x \to B_5} p(x,3), \lim_{x \to B_6} p(x,3), p(c,3)\}$. The curve of $p(x,3)$ is symmetric to $p(x,6)$ around 0. In Theorem~\ref{thm:constraint1}, we will use this property to get compact privacy constraints.
    \begin{figure}[h]
    \centering
      \includegraphics[width=0.85\linewidth]{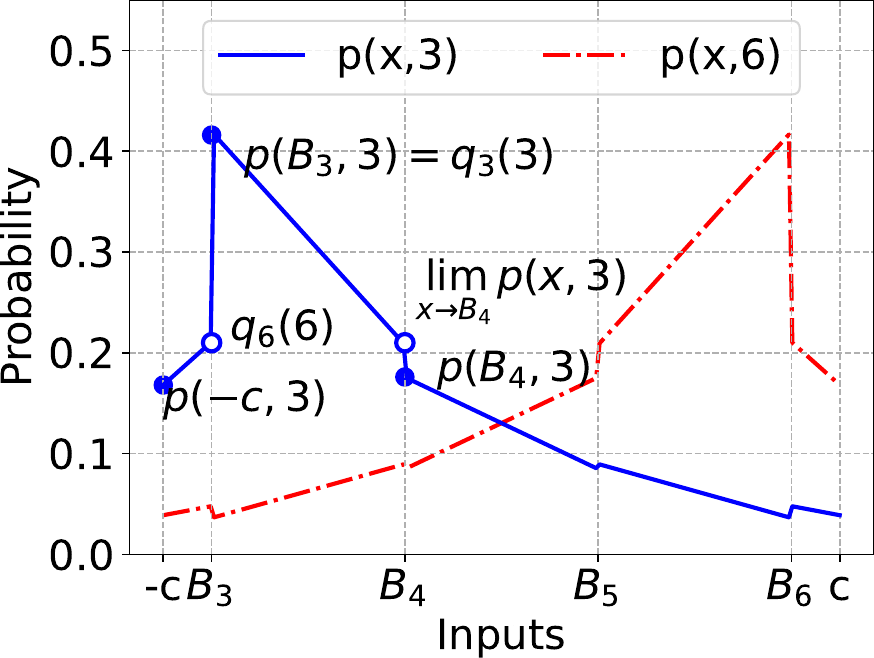}
    \caption{An example of output distribution}
    \label{fig:erm_dist}
\end{figure}
\end{example}


However, $\lim_{x \to B_k}p(x,i)$ and $p(x,i)$ are quadratic functions in $q_j(i)$ (see Section~\ref{subsec:measure}). We still need to convert them into linear forms. To this end, we further assume that each probability $q_j(i)$ has a non-zero lower and upper bound, i.e., $$o_j(i) \leq q_j(i) \leq u_j(i), ~~i,j \in [m], i \leq j,$$ 
where $o_j(i)$ and $u_j(i)$ are hyperparameters that can be found by a grid search. Then, we can replace $q_j(i)$ with $o_j(i)$ or $u_j(i)$ to get a lower bound  $w(x,i)$ or upper bound $z(x,i)$ of $p(x,i)$. We illustrate this using an example.
\begin{example}\label{example}
Consider the following constraint 
\begin{equation}\label{equ:example_constraint}
    p(B_k,i) \leq e^\epsilon \cdot \lim_{x \to B_k}p(x,i).
\end{equation}
where $B_i < B_k$, $p(B_k,i)\in \overline{\mathcal{S}}_i$ and $\lim_{x \to B_k}p(x,i)\in \underline{\mathcal{S}}_i$ by Lemma~\ref{lemma:pr_set}. To make constraint~\eqref{equ:example_constraint} linear, we can replace $p(B_k,i)$ with upper bound $z(B_k,i)$ and replace $\lim_{x \to B_k}p(x,i)$ with a lower bound $w(B_k,i)$. Specifically, $\forall x \in [B_j, B_{j+1})$,  
\begin{equation}\label{equ:def_z}
\resizebox{0.4\textwidth}{!}{$
\displaystyle
    z(x,i) = u_j(i)  \sum_{r \in [j+1,m]} \left( q_{m-j}(m-r+1) \frac{B_r-x}{B_r-B_i}  \right) $}
\end{equation}
\begin{equation}\label{equ:def_w}
 \resizebox{0.48\textwidth}{!}{$
\displaystyle   w(x,i) = 
\begin{cases}
  \displaystyle  o_{j}(i)  \sum_{r \in [j+1, m]}\bigg( q_{m-j}(m-r+1) \frac{B_r-x}{B_r-B_i} \bigg),& \text{ if }x = -c \text{ or } c.\\
 \displaystyle   o_{j-1}(i)  \sum_{r \in [j+1, m]}\bigg( q_{m-j+1}(m-r+1) \frac{B_r-B_j}{B_r-B_{i}} \bigg), & \text{ o.w. }
\end{cases}
$}
\end{equation}

Instead of using constraint~\eqref{equ:example_constraint}, we use a stricter version
\begin{equation*}
z(B_k,i) \leq e^\epsilon \cdot w(B_k,i).
\end{equation*}
\end{example}
Similar to Example~\ref{example}, for any $\overline{s} \in\overline{\mathcal{S}}_i$  and $\underline{s} \in\underline{\mathcal{S}}_i$, we can turn non-linear constraint $\overline{s} \leq e^\epsilon \cdot \underline{s}$ into a stricter version that is linear. Besides, since the bins and the output distribution are symmetric around 0, we only need to find $\min_x p(x,i)$ from $x \in [B_i, c]$ instead of $[-c,c]$. This results in the constraints detailed in Theorem~\ref{thm:constraint1}. 





\begin{theorem}\label{thm:constraint1}
If the bins are symmetric, i.e., $B_{i} = -B_{m+1-i}$, then privacy constraint \eqref{equ:privacy_constraint} can be satisfied if the following $\mathcal{O}(m^3)$ linear constraints are satisfied.
\begin{itemize}[leftmargin=*]
{\small
    \item $\forall i,k \in [m], -c \leq B_{i} < B_k\leq c:$
    \begin{align*}
    & q_i(i) \leq e^\epsilon \cdot w(B_k,i); & & q_{m+1-j}(m+1-j) \leq e^\epsilon \cdot w(B_k,i); \\
    &  q_i(i) \leq e^\epsilon \cdot w(c,i); & & q_{m+1-j}(m+1-j) \leq e^\epsilon \cdot w(c,i);
    \end{align*}
    \item 
$\forall i,k \in [m], B_{i} < -c < B_k:$
\begin{align*}
    z(B_k,i) \leq e^\epsilon \cdot w(B_k,i); & &z(B_k,i) \leq e^\epsilon \cdot w(c,i);\\
    z(-c,i) \leq e^\epsilon \cdot w(B_k,i); &&z(-c,i) \leq e^\epsilon \cdot w(c,i);
\end{align*}
    \item $\forall i,j \in [m], i \leq j, B_{j} \leq c, B_{j+1} > -c:$
    \begin{align*}
     o_{j}(i) \leq q_{j}(i) \leq u_j(i); \qquad 
     q_i(i) \geq q_j(i)
    \end{align*}
    }
\end{itemize}





where $w(\cdot,\cdot)$ and $z(\cdot,\cdot)$ are specified in \eqref{equ:def_w} and \eqref{equ:def_z}. 
\end{theorem}

\paragraph{Complete optimization.} Combining the above results, we can formulate a linear program for the optimal mechanism. Specifically, we minimize the upper bound of MAE in Theorem \ref{thm:uniform} subject to 1) a set of linear constraints in Theorem \ref{thm:constraint1}, and 2) constraint for distribution $q_j$, i.e.,
    $0 \leq q_j(i) \leq 1,
    \sum_{i=1}^j q_j(i)=1, \forall i, j$.
    
The complete procedure for finding the optimal mechanism is shown in Algorithm~\ref{alg:optm}. This optimization can be solved by a linear programming tool denoted by $\mathrm{LinProg}()$, which takes the bin values $\{B_{i}\}_{i\in[m]}$, privacy parameter $ \epsilon$, lower and upper bounds $o_{j}(i), u_{j}(i)$, $i,j\in [m], i\leq j$ as inputs and returns the optimal selection distribution ${q_j(i)}$.

Here we regard $o_{j}(i), u_{j}(i)$ as hyperparameters and use grid search to find the optimal ones. Although bin values $\{B_i\}$ are treated as inputs in Algorithm~\ref{alg:optm}, we can find the optimal bins $\{B_i\}$ using techniques such as grid search to further minimize MAE under a fixed privacy parameter $\epsilon$.  


\paragraph{Reduce complexity.}
As the number of bins $m$ increases, both the number of $q_j(i)$ and the choice of lower and upper bounds $o_j(i), u_j(i)$ increase. Since the optimal $o_j(i), u_j(i)$ are found via grid search, running Algorithm~\ref{alg:optm} can be computationally expensive when $m$ is large. Nonetheless, we can formulate the original privacy constraint \eqref{equ:privacy_constraint} as another set of linear constraints, which are also stricter but significantly reduce the number of $o_j(i)$ and $u_j(i)$ required to conduct the grid search compared to constraints in Theorem~\ref{thm:constraint1}. 

\begin{algorithm}
\caption{\textsf{OPTM}: find optimal selection distribution}
    \label{alg:optm}
    \begin{algorithmic}[1]
        \STATE {\bfseries Input:} bin values $\{B_1,\cdots,B_m\}$, privacy parameter $ \epsilon$
        \STATE $min\_value = \infty$;
        \STATE $P = \emptyset$;
        \FOR{all possible $o_j(i), u_j(i)$ pairs in grid search}
        \STATE $obj, \{q_j(i)\} \gets \mathrm{LinProg}\left(\{B_{i}\}_{i\in[m]}, \epsilon, o_j(i), u_j(i)\right)$;
        
        \IF{$obj \leq min\_value$}
        \STATE $min\_value \gets obj$;
        \STATE $P \gets \{q_j(i)\}$;
        \ENDIF
        \ENDFOR
        \STATE {\bfseries Return:} selection probabilities $P$
    \end{algorithmic}
\end{algorithm}

\begin{theorem}\label{thm:constraint2}
If the bins are symmetric, i.e., $B_{i} = -B_{m+1-i}$, then the privacy constraint \eqref{equ:privacy_constraint} can be satisfied if the following linear constraints are satisfied.
\begin{itemize}[leftmargin=*]
{\small 
    \item $\forall i,j \in [m], i \leq j:$
        \begin{align*}
            q_j(i) \geq q_{j+1}(i); && q_j(i) \leq q_{j}(i+1);
        \end{align*}
    \item $\forall i\in [m], -c \leq B(i) \leq c:$
        \begin{align*}
            q_i(i) \geq q_{i+1}(i+1);
        \end{align*}
    \item Let $B_s<B_t$ be bins in $[-c,c]$ closest to $-c$ and $c$, respectively:
  \begin{align*}
           & z(-c,s-1) \leq e^{\epsilon} \cdot w(B_t,1); & & q_s(s) \leq e^{\epsilon} \cdot w(B_t,1);  \\
           & z(-c,s-1) \leq e^{\epsilon} \cdot w(c,1); & & q_s(s) \leq e^{\epsilon} \cdot w(c,1).
        \end{align*}
    \item $\forall r, k \in [m], s \leq k \leq t, r > k+1$:
        \begin{align*}
            \frac{q_{m-k+1}(m-r+1)}{B_{r}-B_{k+1}}& \geq &\frac{q_{m-k}(m-r+1)}{B_{r}-B_{k}}; \\
            \frac{q_{m-k}(m-r+1)}{B_{r}-B_{k+1}}& \geq &\frac{q_{m-k-1}(m-r+1)}{B_{r}-B_{k}};
        \end{align*}
        }
\end{itemize}

\end{theorem}
The constraints in Theorem~\ref{thm:constraint2} induces that
\begin{align*}
& \max_{x,i} p(x,i) \in \left\{p(-c,s-1), q_s(s)\right\}; \\
& \min_{x,i} p(x,i) \in \left\{\lim_{x \to B_t} p(x,1), p(c,1)\right\},
\end{align*}
where $s$ and $t$ are as defined in Theorem~\ref{thm:constraint2}. Under this set of linear constraints, the upper bound $u_j(i)$ only appears when calculating $z(-c,s-1)$, and the lower bound $o_j(i)$ is only used for computing $w(B_t,1)$ and $w(c,1)$. Thus, we can conduct a grid search over 3 variables, regardless of the number of output bits. 


\section{Discussion}
Our proposed mechanisms can be generalized to broader settings, including dynamic, high-dimensional, and biased quantization. We discuss these extensions below. 

\paragraph{Extension to high-dimensional quantization.}
Besides entry-wise discretization, our method can also be extended to higher-dimensional quantization with a similar method as in~\citep{mvu}. Specifically, for any $d$-dimensional input vector $\textbf{x}=(\textbf{x}_1, \cdots, \textbf{x}_d)$ with $L_2$ norm bounded by diameter $B$, we map the input vector $\textbf{x}$ to $\mathcal{M}_d(\textbf{x})=(\mathcal{M}^{\prime}(\textbf{x}_1), \cdots, \mathcal{M}^{\prime}(\textbf{x}_d))$. Here, the mechanism $\mathcal{M}^{\prime}$ quantize the scalar in each coordinate and needs to satisfy $\epsilon$-metric DP, a variant of $\epsilon$-DP that requires the following holds for any two inputs $x, x^{\prime}$ and any set of possible outputs $S \subseteq$ Range($\mathcal{M}$): $$\Pr(\mathcal{M}^{\prime}(x) \in S) \leq e^{\epsilon d(x, x^{\prime})} \Pr(\mathcal{M}^{\prime}(x^{\prime}) \in S),$$ where $d(x, x^{\prime})=|x - x^{\prime}|^2$. Since Lemma 6 in~\cite{mvu} has shown that the mechanism $\mathcal{M}_d$ generated by $\epsilon$-metric DP $\mathcal{M}^{\prime}$ is $\epsilon B^2$-DP and unbiased, we can directly use our method to find the optimal parameters of $\mathcal{M}^{\prime}$ (under new privacy constraints specified by $\epsilon$-metric DP).

\paragraph{Extension to biased quantization.}
Our unbiased mechanism can be extended to biased quantization, finding a new tradeoff between bias, deviation, and privacy. Instead of randomly outputting either $B_l$ or $B_r$ and enforcing unbiasedness according to Eq.~\eqref{equ:dither} as defined in Section~\ref{subsec:mechanism}, we can use the exponential mechanism to output either $B_l$ or $B_r$, with score function being the negative distance between the input and output bins. This mechanism induces biased output but reduces privacy loss. 

\paragraph{Extension to dynamic settings.} 
Our method can also be extended to dynamic quantization, where different inputs require quantization mechanisms with different hyperparameters. One potential solution is to integrate the existing dynamic quantization strategies, such as the optimal quantization bit-width~\citep{dyn_quan_bit}, the clipping range of activation values~\citep{dyn_quan_act} in a quantized neural network; both methods find hyperparameters (e.g., number of bins, clipping range) during runtime. After these hyperparameters are decided and samples of inputs are collected, we can directly use our algorithm to find the optimal quantization mechanism. 

\section{Experiments}\label{sec:res}

Next, we validate two proposed mechanisms: 1) optimal randomized quantization mechanism (\textsf{OPTM}) proposed in Section~\ref{subsec:gen}; 2) exponential randomized
mechanism (\textsf{ERM}), a special case of \textsf{OPTM} proposed in Section~\ref{subsec:special}. We use grid search to find bin values with the best performance.

We conduct three sets of experiments: (i) scalar input quantization; (ii) vector input quantization; and (iii) quantization in stochastic gradient descent (SGD). For each experiment, we compare our mechanisms  with two baselines: 
\begin{itemize}[leftmargin=*]
    \item Randomized quantization mechanism (\textsf{RQM}) \citep{rqm}:  a special case of \textsf{OPTM} with uniformly-distributed bins as discussed in Section~\ref{subsec:special}.   \item Minimum variance
unbiased (\textsf{MVU}) mechanism \citep{mvu}: a mechanism that uses optimized probability matrix and output alphabets to map the quantized inputs to outputs. It finds the optimal bin values via a non-linear optimization.
\end{itemize}
For each mechanism, we evaluate the privacy and accuracy using the standard differential privacy (DP) and mean absolute error (MAE) measures. 

\begin{figure*}[htbp]

\begin{subfigure}{0.33\textwidth}
\centering
\includegraphics[width=\linewidth]{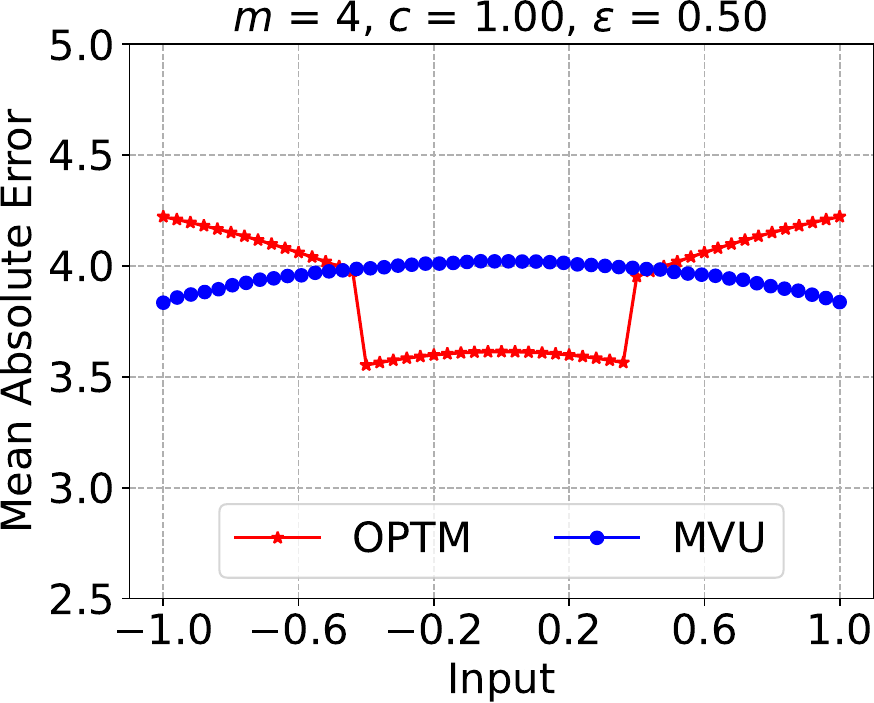} 
\end{subfigure}
\begin{subfigure}{0.33\textwidth}
\includegraphics[width=\linewidth]{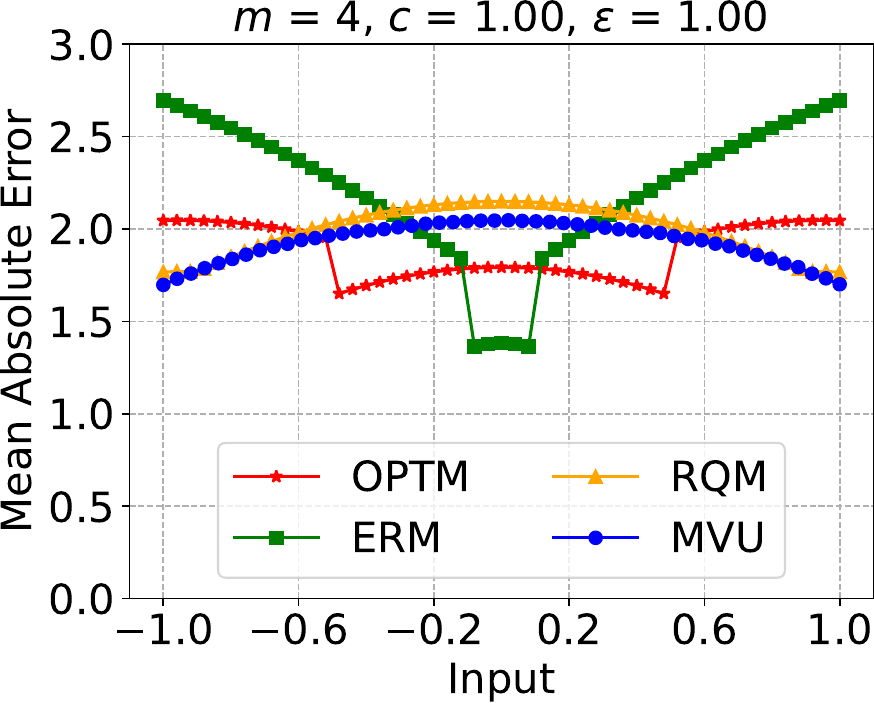}
\end{subfigure}
\begin{subfigure}{0.33\textwidth}
\includegraphics[width=\linewidth]{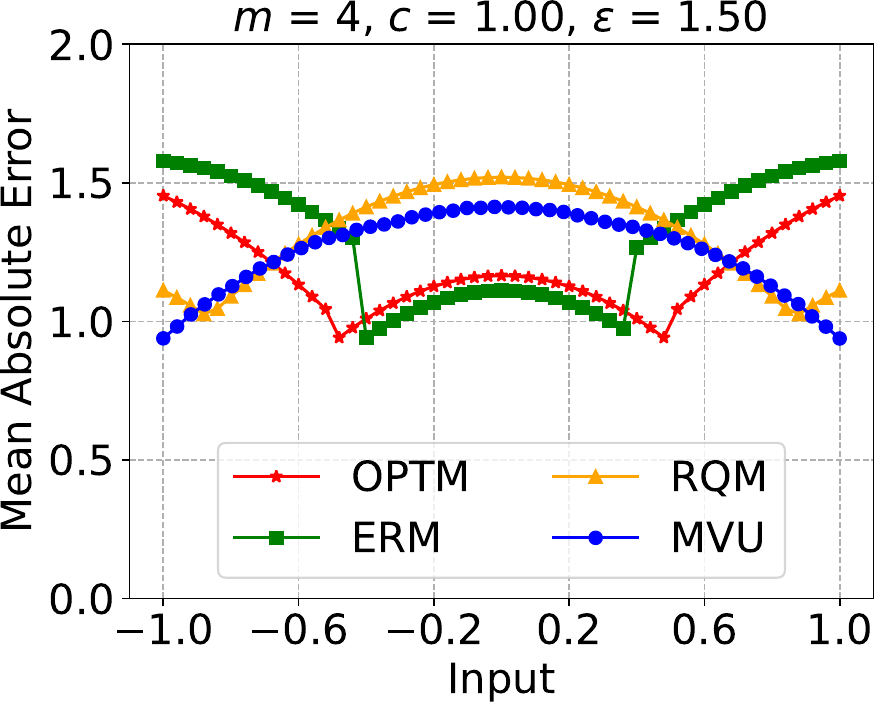}
\end{subfigure}
\caption{Comparison of mean absolute error under the same privacy on scalar inputs}
\label{fig:exp_err}
\end{figure*}

\begin{figure*}[ht]
\begin{subfigure}{0.24\textwidth}
\centering
\includegraphics[width=\linewidth]{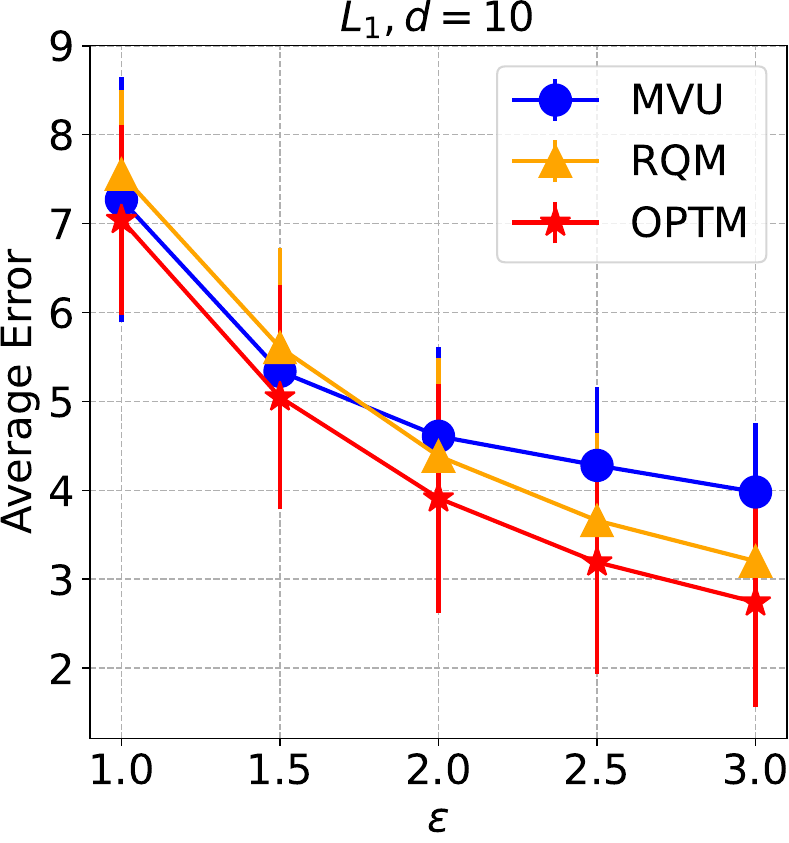} 
\caption{}\label{fig:l1_vec_err}
\end{subfigure}
\begin{subfigure}{0.247\textwidth}
\includegraphics[width=\linewidth]{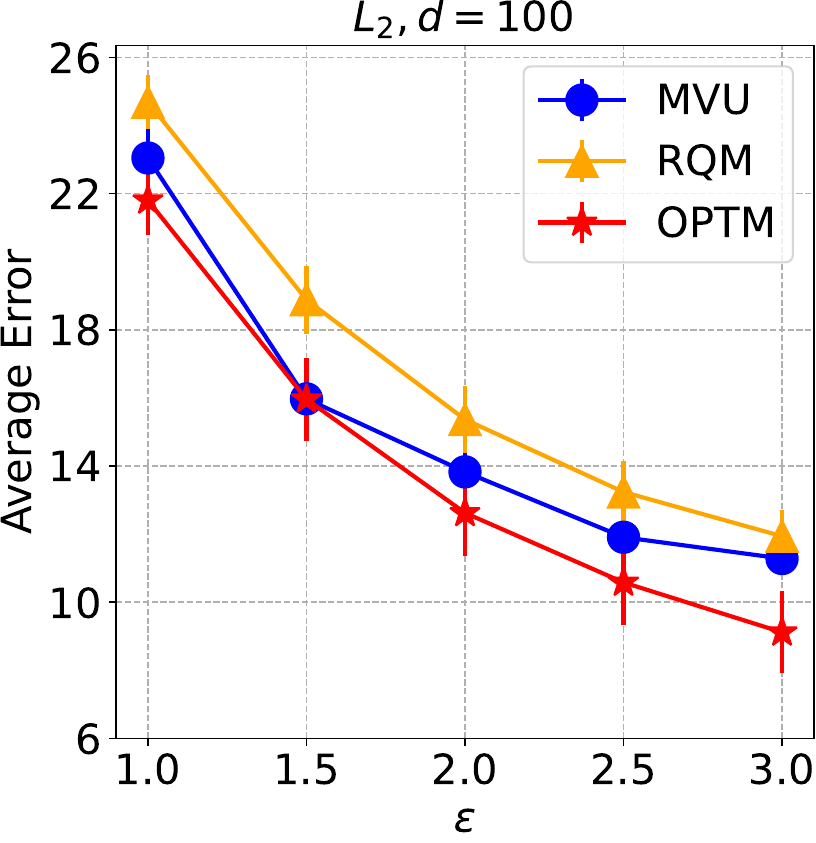}\caption{}
\label{fig:l2_vec_err}
\end{subfigure}
\begin{subfigure}{0.25\textwidth}
\centering
\includegraphics[width=\linewidth]{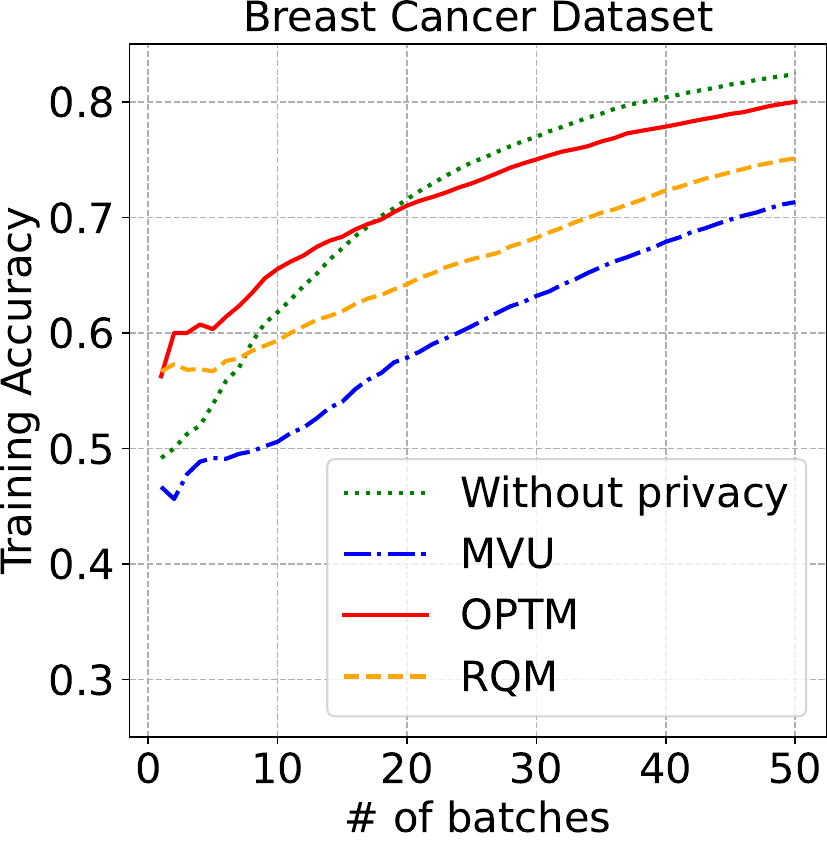} \caption{}\label{fig:acc_dp_sgd1}
\end{subfigure}
\begin{subfigure}{0.25\textwidth}
\includegraphics[width=\linewidth]{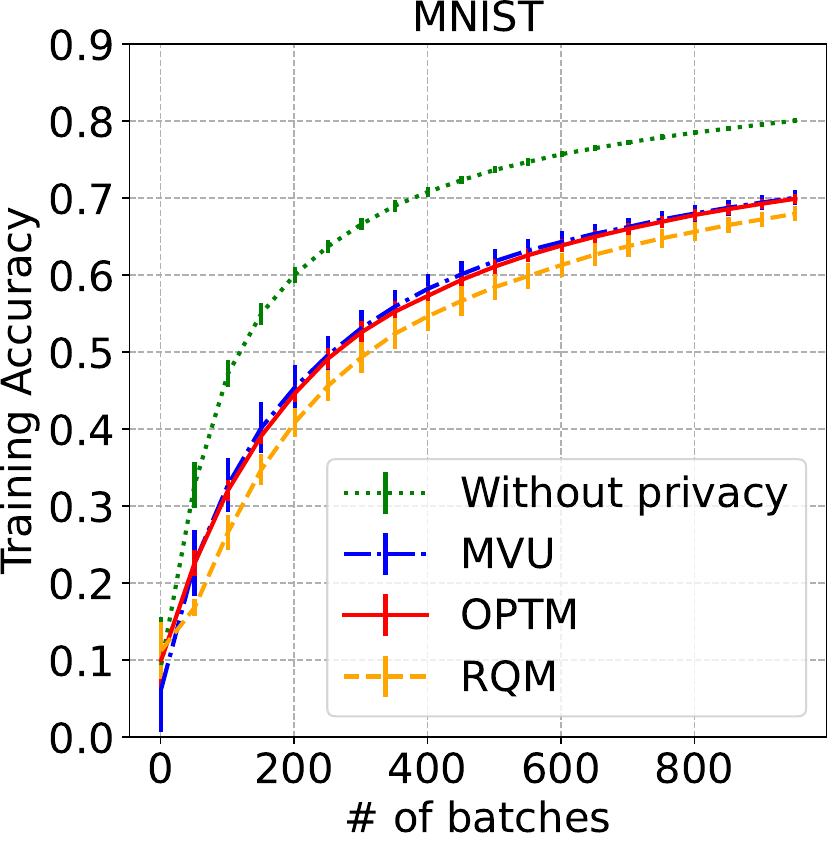}\caption{}\label{fig:acc_dp_sgd2}

\end{subfigure}
\caption{a) Average error of $L_1$ bounded vectors , b) Average error of $L_2$ bounded vectors, c) Training accuracy on breast cancer dataset, d) Training accuracy on MNIST dataset}
\label{fig:acc_dp_sgd}
\end{figure*}

\subsection{Scalar input} We first evaluate the performance of our algorithm and baselines on a scalar input. In our experiments, the time and resources needed to find the optimal parameters for \textsf{OPTM} are low. It takes about 300 seconds on a personal computer (with Intel Core i5-10210U CPU and 16 GB RAM) to search over all combinations of hyperparameters (10 optional $\Delta$, 10 optional bin assignments, 100 optional lower/upper bounds on probabilities), and find the parameters which can induce the best performance.

We first consider a scenario where input $x$ follows a uniform distribution over $[-1,1]$. Table \ref{tab:min_avg_err} compares the mean absolute error $\mathbb{E}_{X}(|\mathcal{M}(X)-X|)$ at $\epsilon = 0.5,1.0,1.5$ when $m=4$. As expected, our method \textsf{OPTM} improves privacy-accuracy trade-off, and it has the lowest error compared to baselines. The performance of \textsf{ERM} is also comparable with \textsf{RQM}. It is worth mentioning that we could not find valid hyperparameters for \textsf{RQM} and \textsf{ERM} when privacy loss $\epsilon=0.5$ so we put "N/A" in Table~\ref{tab:min_avg_err}. Figure \ref{fig:exp_err} illustrates mean absolute error with higher granularity for each input value $x$.
The choice of parameters in each mechanism are given in the Appendix~\ref{exp_detail}. We scale the input range of \textsf{MVU} to $[-1,1]$ for a fair comparison and also scale the output alphabets. The results show that \textsf{ERM} can achieve similar and sometimes better utility than \textsf{RQM}. \textsf{OPTM} can achieve lower error in most cases compared to \textsf{RQM} and \textsf{ERM}, which indicates the effectiveness of the optimization scheme. 

We then consider a scenario where input $x$ follows a truncated Gaussian distribution and the distribution is not known in advance. Specifically, the input is first sampled from Gaussian distribution with $\mu = 0.5$, $\sigma = 0.1, 0.2, 0.3$, and then truncated by $[-1, 1]$. Table \ref{tab:min_avg_err_gauss} compares the mean absolute error when $m=4$ and $\epsilon=1$. The results show that \textsf{OPTM} can use asymmetric bins to better capture the pattern of the underlying distribution. Specifically, for each optional bin value, we use samples collected from the same input distribution to estimate the density function, optimize for the parameters with the objective function as stated in Theorem~\ref{thm:extend}, and find the bin values inducing the best performance. In comparison, \textsf{MVU} and \textsf{RQM} use uniformly distributed bins for all inputs, hence inducing higher errors. The performance gain brought by asymmetric bins is higher when the distribution is more concentrated (i.e., with smaller $\sigma$).

\begin{table}[h]
    \caption{Minimal MAE of scalar inputs under uniform distribution. \textsf{OPTM} attains higher accuracy than baselines. N/A means that there are no valid hyperparameters for \textsf{ERM} and \textsf{RQM} when $\epsilon=0.5$.  }
\label{tab:min_avg_err}
    \centering
    \begin{tabular}{cccc}
    \toprule
    $ \mathbb{E}_{X}(|\mathcal{M}(X)-X|)$ & $\epsilon=0.5$ & $\epsilon=1$ & $\epsilon=1.5$ \\
    \midrule
    \textsf{OPTM} & 3.904 & 1.882 & 1.179 \\
    \textsf{MVU} & 3.959 & 1.930 & 1.254 \\
    \textsf{RQM} & N/A & 1.993 & 1.310 \\
    \textsf{ERM} & N/A & 2.216 & 1.304 \\
   \bottomrule
\end{tabular}
\end{table}

\begin{table}[h]
    \caption{Minimal MAE of scalar inputs under truncated Gaussian distribution. Our proposed \textsf{OPTM} attains higher accuracy than baselines.}
\label{tab:min_avg_err_gauss}
    \centering
    \begin{tabular}{cccc}
    \toprule
    $ \mathbb{E}_{X}(|\mathcal{M}(X)-X|)$ & $\sigma=0.1$ & $\sigma=0.2$ & $\sigma=0.3$ \\
    \midrule
    \textsf{OPTM} & 1.778 & 1.836 & 1.972 \\
    \textsf{MVU} & 2.053 & 2.052 & 2.002 \\
    \textsf{RQM} & 2.028 & 2.010 & 2.000 \\
   \bottomrule
\end{tabular}
\vspace{-0.1cm}
\end{table}

\subsection{Vector input} We then compare the error of our mechanism with vector inputs under privacy parameter $\epsilon$. Hyperparameters of each mechanism are given in the Appendix~\ref{exp_detail}. We use bounded random vectors as inputs to simulate the clipped gradients in DP-SGD~\citep{dp-sgd}, i.e., differentially private stochastic gradient descent commonly used for training private machine learning models. Specifically, we generate random vectors with dimension $d=10$. Each coordinate follows uniform distribution in $[-1,1]$, hence producing vectors with bounded $L_1$ norm. 

For each $\epsilon$, we fix bin values $\{B_1,\ldots,B_m\}$ and find the optimal parameters for each mechanism (e.g., selection probability in \textsf{OPTM} and parameter $q$ for \textsf{RQM}). Then, we quantize each coordinate independently.  
 We measure the Euclidean distance between the input and output vector as the error, and repeat this process 10,000 times to calculate the average error (see Figure \ref{fig:l1_vec_err}). 
 
 In another experiment (Figure \ref{fig:l2_vec_err}), we generate random vectors $v\in \mathbb{R}^{100}$ with uniform distribution over ball $||v||_2\leq 1$ (this can be done  through ball point picking~\citep{ball}). We quantize the vector $v$ and measure the error based on Euclidean distance. Again we repeat the process 10,000 times to find the average error. We report both the mean and the standard deviation of the error 
 in Figure \ref{fig:l1_vec_err} and  \ref{fig:l2_vec_err}. In both cases, \textsf{OPTM} can achieve lower error compared to \textsf{RQM} and \textsf{MVU}, indicating that our mechanism can effectively reduce the loss when privatizing vector inputs.

\subsection{DP Stochastic Gradient Descent } We further measure the performance of our mechanisms on downstream machine learning tasks by integrating them into DP-SGD~\citep{dp-sgd} algorithms. Specifically, during each epoch of the Stochastic Gradient Descent (SGD), each coordinate of the gradient vector is clipped by a threshold and then quantized by 
differentially private mechanisms. The parameters of the experiments are given in the Appendix~\ref{exp_detail}. We also record the accuracy when gradients are only clipped, without any privacy protection.  

In our experiments, we first use DP-SGD to train a softmax regression model based on the UCI ML Breast Cancer 
dataset~\citep{cancer} with 569 samples. We record the accuracy on the training set after training on each batch of data. Results are shown in Figure~\ref{fig:acc_dp_sgd1}. We also train a softmax regression model based on the MNIST dataset~\citep{mnist} with 60,000 images and record the training accuracy. Results are shown in Figure~\ref{fig:acc_dp_sgd2}.  
On the Breast Cancer dataset, \textsf{OPTM} achieves a better convergence rate than \textsf{RQM} and \textsf{MVU}, and achieves very close accuracy compared with the non-private scheme. 
On MNIST dataset, \textsf{OPTM} has the same performance as \textsf{MVU} and higher accuracy compared to \textsf{RQM}. As errors brought by DP mechanisms can slow down the convergence process, our mechanism can achieve a better convergence rate compared to baselines.

\section{Conclusion}\label{sec:con}

This paper proposes a family of differential privacy mechanisms with discrete and unbiased outputs, which is desirable in many real applications such as federated learning. 
We design an efficient linear programming algorithm to find the optimal parameters for our mechanism. Experiments on synthetic and real data show that the proposed mechanisms can achieve a better accuracy-privacy trade-off compared with existing discrete differential privacy mechanisms.

Our research raises several interesting topics for future research: (i) Finding the optimal hyperparameters (e.g., number of bins, clipping range) automatically with an optimization during runtime. In this paper, we assume that these hyperparameters are either given in advance or are found using grid search with the best performance. (ii) Privacy analysis of \textsf{OPTM}. This paper only quantifies the privacy loss of exponential randomized mechanism (\textsf{ERM}) in Theorem \ref{thm:erm_privacy}, finding the privacy loss for the optimal \textsf{OPTM} is important and allows us to better understand the relationship between the privacy bound and mechanism parameters.

\section{Acknowledgement}
This material is based upon work supported by the U.S. National Science Foundation under award IIS-2202699, IIS-2301599, and ECCS-2301601, by two grants from the Ohio State University Translational Data Analytics
Institute, and the College of Engineering Strategic Research Initiative Grant at the Ohio State University.

\bibliography{main}

\begin{thebibliography}{40}
\providecommand{\natexlab}[1]{#1}
\providecommand{\url}[1]{\texttt{#1}}
\expandafter\ifx\csname urlstyle\endcsname\relax
  \providecommand{\doi}[1]{doi: #1}\else
  \providecommand{\doi}{doi: \begingroup \urlstyle{rm}\Url}\fi

\bibitem[Abadi et~al.(2016)Abadi, Chu, Goodfellow, McMahan, Mironov, Talwar, and Zhang]{dp-sgd}
Martin Abadi, Andy Chu, Ian Goodfellow, H.~Brendan McMahan, Ilya Mironov, Kunal Talwar, and Li~Zhang.
\newblock Deep learning with differential privacy.
\newblock In \emph{Proceedings of the 2016 ACM SIGSAC Conference on Computer and Communications Security}, CCS '16, page 308–318, New York, NY, USA, 2016. Association for Computing Machinery.
\newblock ISBN 9781450341394.
\newblock \doi{10.1145/2976749.2978318}.
\newblock URL \url{https://doi.org/10.1145/2976749.2978318}.

\bibitem[Abowd(2018)]{abowd}
John~M Abowd.
\newblock Disclosure avoidance for block level data and protection of confidentiality in public tabulations.
\newblock In \emph{Census Scientific Advisory Committee (Fall Meeting)}, pages 2018--12, 2018.

\bibitem[Agarwal et~al.(2021)Agarwal, Kairouz, and Liu]{skellam}
Naman Agarwal, Peter Kairouz, and Ziyu Liu.
\newblock The skellam mechanism for differentially private federated learning.
\newblock \emph{Advances in Neural Information Processing Systems}, 34:\penalty0 5052--5064, 2021.

\bibitem[Bai et~al.(2022)Bai, Hou, Shang, Jiang, King, and Lyu]{post-train-quant}
Haoli Bai, Lu~Hou, Lifeng Shang, Xin Jiang, Irwin King, and Michael~R Lyu.
\newblock Towards efficient post-training quantization of pre-trained language models.
\newblock \emph{Advances in Neural Information Processing Systems}, 35:\penalty0 1405--1418, 2022.

\bibitem[Barthe et~al.(2005)Barthe, Gu{\'e}don, Mendelson, and Naor]{ball}
Franck Barthe, Olivier Gu{\'e}don, Shahar Mendelson, and Assaf Naor.
\newblock A probabilistic approach to the geometry of the $\ell^n_p$-ball.
\newblock 2005.

\bibitem[Bassily et~al.(2016)Bassily, Nissim, Smith, Steinke, Stemmer, and Ullman]{adaptive}
Raef Bassily, Kobbi Nissim, Adam Smith, Thomas Steinke, Uri Stemmer, and Jonathan Ullman.
\newblock Algorithmic stability for adaptive data analysis.
\newblock In \emph{Proceedings of the forty-eighth annual ACM symposium on Theory of Computing}, pages 1046--1059, 2016.

\bibitem[Bonawitz et~al.(2017)Bonawitz, Ivanov, Kreuter, Marcedone, McMahan, Patel, Ramage, Segal, and Seth]{secagg}
Keith Bonawitz, Vladimir Ivanov, Ben Kreuter, Antonio Marcedone, H.~Brendan McMahan, Sarvar Patel, Daniel Ramage, Aaron Segal, and Karn Seth.
\newblock Practical secure aggregation for privacy-preserving machine learning.
\newblock In \emph{Proceedings of the 2017 ACM SIGSAC Conference on Computer and Communications Security}, CCS '17, page 1175–1191, New York, NY, USA, 2017. Association for Computing Machinery.
\newblock ISBN 9781450349468.
\newblock \doi{10.1145/3133956.3133982}.
\newblock URL \url{https://doi.org/10.1145/3133956.3133982}.

\bibitem[Bottou et~al.(2018)Bottou, Curtis, and Nocedal]{optimization}
L{\'e}on Bottou, Frank~E Curtis, and Jorge Nocedal.
\newblock Optimization methods for large-scale machine learning.
\newblock \emph{SIAM review}, 60\penalty0 (2):\penalty0 223--311, 2018.

\bibitem[Canonne et~al.(2020)Canonne, Kamath, and Steinke]{discrete-gaussian}
Cl{\'e}ment~L Canonne, Gautam Kamath, and Thomas Steinke.
\newblock The discrete gaussian for differential privacy.
\newblock \emph{Advances in Neural Information Processing Systems}, 33:\penalty0 15676--15688, 2020.

\bibitem[Chaudhuri et~al.(2022)Chaudhuri, Guo, and Rabbat]{mvu}
Kamalika Chaudhuri, Chuan Guo, and Mike Rabbat.
\newblock Privacy-aware compression for federated data analysis.
\newblock In James Cussens and Kun Zhang, editors, \emph{Proceedings of the Thirty-Eighth Conference on Uncertainty in Artificial Intelligence}, volume 180 of \emph{Proceedings of Machine Learning Research}, pages 296--306. PMLR, 01--05 Aug 2022.
\newblock URL \url{https://proceedings.mlr.press/v180/chaudhuri22a.html}.

\bibitem[Chen et~al.(2022)Chen, Ozgur, and Kairouz]{pbm}
Wei-Ning Chen, Ayfer Ozgur, and Peter Kairouz.
\newblock The poisson binomial mechanism for unbiased federated learning with secure aggregation.
\newblock In \emph{International Conference on Machine Learning}, pages 3490--3506. PMLR, 2022.

\bibitem[Dwork(2006)]{dwork2006differential}
Cynthia Dwork.
\newblock Differential privacy.
\newblock In \emph{International Colloquium on Automata, Languages, and Programming}, pages 1--12. Springer, 2006.

\bibitem[Dwork et~al.(2006)Dwork, McSherry, Nissim, and Smith]{laplace}
Cynthia Dwork, Frank McSherry, Kobbi Nissim, and Adam Smith.
\newblock Calibrating noise to sensitivity in private data analysis.
\newblock In \emph{Proceedings of the Third Conference on Theory of Cryptography}, TCC'06, page 265–284, Berlin, Heidelberg, 2006. Springer-Verlag.
\newblock ISBN 3540327312.
\newblock \doi{10.1007/11681878_14}.
\newblock URL \url{https://doi.org/10.1007/11681878_14}.

\bibitem[Dwork et~al.(2014)Dwork, Roth, et~al.]{gaussian}
Cynthia Dwork, Aaron Roth, et~al.
\newblock The algorithmic foundations of differential privacy.
\newblock \emph{Foundations and Trends{\textregistered} in Theoretical Computer Science}, 9\penalty0 (3--4):\penalty0 211--407, 2014.

\bibitem[Gandikota et~al.(2021)Gandikota, Kane, Maity, and Mazumdar]{vqsgd}
Venkata Gandikota, Daniel Kane, Raj~Kumar Maity, and Arya Mazumdar.
\newblock vqsgd: Vector quantized stochastic gradient descent.
\newblock In \emph{International Conference on Artificial Intelligence and Statistics}, pages 2197--2205. PMLR, 2021.

\bibitem[Ghosh et~al.(2009)Ghosh, Roughgarden, and Sundararajan]{discrete-laplace}
Arpita Ghosh, Tim Roughgarden, and Mukund Sundararajan.
\newblock Universally utility-maximizing privacy mechanisms.
\newblock In \emph{Proceedings of the Forty-First Annual ACM Symposium on Theory of Computing}, STOC '09, page 351–360, New York, NY, USA, 2009. Association for Computing Machinery.
\newblock ISBN 9781605585062.
\newblock \doi{10.1145/1536414.1536464}.
\newblock URL \url{https://doi.org/10.1145/1536414.1536464}.

\bibitem[Girgis et~al.(2021)Girgis, Data, Diggavi, Kairouz, and Theertha~Suresh]{shuffled}
Antonious Girgis, Deepesh Data, Suhas Diggavi, Peter Kairouz, and Ananda Theertha~Suresh.
\newblock Shuffled model of differential privacy in federated learning.
\newblock In Arindam Banerjee and Kenji Fukumizu, editors, \emph{Proceedings of The 24th International Conference on Artificial Intelligence and Statistics}, volume 130 of \emph{Proceedings of Machine Learning Research}, pages 2521--2529. PMLR, 13--15 Apr 2021.
\newblock URL \url{https://proceedings.mlr.press/v130/girgis21a.html}.

\bibitem[Guo et~al.(2023)Guo, Chaudhuri, Stock, and Rabbat]{i-mvu}
Chuan Guo, Kamalika Chaudhuri, Pierre Stock, and Mike Rabbat.
\newblock Privacy-aware compression for federated learning through numerical mechanism design, 2023.

\bibitem[H{\"o}nig et~al.(2022)H{\"o}nig, Zhao, and Mullins]{dadaquant}
Robert H{\"o}nig, Yiren Zhao, and Robert Mullins.
\newblock {DA}da{Q}uant: Doubly-adaptive quantization for communication-efficient federated learning.
\newblock In Kamalika Chaudhuri, Stefanie Jegelka, Le~Song, Csaba Szepesvari, Gang Niu, and Sivan Sabato, editors, \emph{Proceedings of the 39th International Conference on Machine Learning}, volume 162 of \emph{Proceedings of Machine Learning Research}, pages 8852--8866. PMLR, 17--23 Jul 2022.
\newblock URL \url{https://proceedings.mlr.press/v162/honig22a.html}.

\bibitem[Hopkins et~al.(2022)Hopkins, Kamath, and Majid]{hopkins2022efficient}
Samuel~B Hopkins, Gautam Kamath, and Mahbod Majid.
\newblock Efficient mean estimation with pure differential privacy via a sum-of-squares exponential mechanism.
\newblock In \emph{Proceedings of the 54th Annual ACM SIGACT Symposium on Theory of Computing}, pages 1406--1417, 2022.

\bibitem[Jayaraman and Evans(2019)]{jayaraman2019evaluating}
Bargav Jayaraman and David Evans.
\newblock Evaluating differentially private machine learning in practice.
\newblock In \emph{28th USENIX Security Symposium (USENIX Security 19)}, pages 1895--1912, 2019.

\bibitem[Jin et~al.(2024)Jin, Yin, Chen, Sun, Zhang, Liu, and Liu]{jin2024performative}
Kun Jin, Tongxin Yin, Zhongzhu Chen, Zeyu Sun, Xueru Zhang, Yang Liu, and Mingyan Liu.
\newblock Performative federated learning: A solution to model-dependent and heterogeneous distribution shifts.
\newblock In \emph{Proceedings of the AAAI Conference on Artificial Intelligence}, volume~38, pages 12938--12946, 2024.

\bibitem[Kairouz et~al.(2021)Kairouz, Liu, and Steinke]{dis_gau_fed}
Peter Kairouz, Ziyu Liu, and Thomas Steinke.
\newblock The distributed discrete gaussian mechanism for federated learning with secure aggregation.
\newblock In Marina Meila and Tong Zhang, editors, \emph{Proceedings of the 38th International Conference on Machine Learning}, volume 139 of \emph{Proceedings of Machine Learning Research}, pages 5201--5212. PMLR, 18--24 Jul 2021.
\newblock URL \url{https://proceedings.mlr.press/v139/kairouz21a.html}.

\bibitem[Khalili et~al.(2021{\natexlab{a}})Khalili, Zhang, Abroshan, and Sojoudi]{khalili2021improving}
Mohammad~Mahdi Khalili, Xueru Zhang, Mahed Abroshan, and Somayeh Sojoudi.
\newblock Improving fairness and privacy in selection problems.
\newblock In \emph{Proceedings of the AAAI Conference on Artificial Intelligence}, volume~35, pages 8092--8100, 2021{\natexlab{a}}.

\bibitem[Khalili et~al.(2021{\natexlab{b}})Khalili, Zhang, and Liu]{khalili2021designing}
Mohammad~Mahdi Khalili, Xueru Zhang, and Mingyan Liu.
\newblock Designing contracts for trading private and heterogeneous data using a biased differentially private algorithm.
\newblock \emph{IEEE Access}, 9:\penalty0 70732--70745, 2021{\natexlab{b}}.

\bibitem[LeCun et~al.(2010)LeCun, Cortes, and Burges]{mnist}
Yann LeCun, Corinna Cortes, and CJ~Burges.
\newblock Mnist handwritten digit database.
\newblock \emph{ATT Labs [Online]. Available: http://yann.lecun.com/exdb/mnist}, 2, 2010.

\bibitem[Liu et~al.(2021)Liu, Kong, Kakade, and Oh]{liu2021robust}
Xiyang Liu, Weihao Kong, Sham Kakade, and Sewoong Oh.
\newblock Robust and differentially private mean estimation.
\newblock \emph{Advances in neural information processing systems}, 34:\penalty0 3887--3901, 2021.

\bibitem[McSherry and Talwar(2007)]{mcsherry2007mechanism}
Frank McSherry and Kunal Talwar.
\newblock Mechanism design via differential privacy.
\newblock In \emph{48th Annual IEEE Symposium on Foundations of Computer Science (FOCS'07)}, pages 94--103. IEEE, 2007.

\bibitem[Mironov(2012)]{least_bit}
Ilya Mironov.
\newblock On significance of the least significant bits for differential privacy.
\newblock In \emph{Proceedings of the 2012 ACM Conference on Computer and Communications Security}, CCS '12, page 650–661, New York, NY, USA, 2012. Association for Computing Machinery.
\newblock ISBN 9781450316514.
\newblock \doi{10.1145/2382196.2382264}.
\newblock URL \url{https://doi.org/10.1145/2382196.2382264}.

\bibitem[Reisizadeh et~al.(2020)Reisizadeh, Mokhtari, Hassani, Jadbabaie, and Pedarsani]{fedpaq}
Amirhossein Reisizadeh, Aryan Mokhtari, Hamed Hassani, Ali Jadbabaie, and Ramtin Pedarsani.
\newblock Fedpaq: A communication-efficient federated learning method with periodic averaging and quantization.
\newblock In Silvia Chiappa and Roberto Calandra, editors, \emph{Proceedings of the Twenty Third International Conference on Artificial Intelligence and Statistics}, volume 108 of \emph{Proceedings of Machine Learning Research}, pages 2021--2031. PMLR, 26--28 Aug 2020.
\newblock URL \url{https://proceedings.mlr.press/v108/reisizadeh20a.html}.

\bibitem[So et~al.(2024)So, Lee, Ahn, Kim, and Park]{dyn_quan_act}
Junhyuk So, Jungwon Lee, Daehyun Ahn, Hyungjun Kim, and Eunhyeok Park.
\newblock Temporal dynamic quantization for diffusion models.
\newblock \emph{Advances in Neural Information Processing Systems}, 36, 2024.

\bibitem[Tao et~al.(2022)Tao, Hou, Zhang, Shang, Jiang, Liu, Luo, and Wong]{compress-model}
Chaofan Tao, Lu~Hou, Wei Zhang, Lifeng Shang, Xin Jiang, Qun Liu, Ping Luo, and Ngai Wong.
\newblock Compression of generative pre-trained language models via quantization.
\newblock \emph{arXiv preprint arXiv:2203.10705}, 2022.

\bibitem[Wang and Zhou(2020)]{wang2020differentially}
Jun Wang and Zhi-Hua Zhou.
\newblock Differentially private learning with small public data.
\newblock In \emph{Proceedings of the AAAI Conference on Artificial Intelligence}, volume~34, pages 6219--6226, 2020.

\bibitem[{Wolberg,William, Mangasarian,Olvi, Street,Nick, and Street,W.}(1995)]{cancer}
{Wolberg,William, Mangasarian,Olvi, Street,Nick, and Street,W.}
\newblock {Breast Cancer Wisconsin (Diagnostic)}.
\newblock UCI Machine Learning Repository, 1995.
\newblock {DOI}: https://doi.org/10.24432/C5DW2B.

\bibitem[Youn et~al.(2023)Youn, Hu, Ziani, and Abernethy]{rqm}
Yeojoon Youn, Zihao Hu, Juba Ziani, and Jacob Abernethy.
\newblock Randomized quantization is all you need for differential privacy in federated learning, 2023.

\bibitem[Zhang et~al.(2018{\natexlab{a}})Zhang, Khalili, and Liu]{zhang2018improving}
Xueru Zhang, Mohammad~Mahdi Khalili, and Mingyan Liu.
\newblock Improving the privacy and accuracy of admm-based distributed algorithms.
\newblock In \emph{International Conference on Machine Learning}, pages 5796--5805. PMLR, 2018{\natexlab{a}}.

\bibitem[Zhang et~al.(2018{\natexlab{b}})Zhang, Khalili, and Liu]{zhang2018recycled}
Xueru Zhang, Mohammad~Mahdi Khalili, and Mingyan Liu.
\newblock Recycled admm: Improve privacy and accuracy with less computation in distributed algorithms.
\newblock In \emph{2018 56th Annual Allerton Conference on Communication, Control, and Computing (Allerton)}, pages 959--965. IEEE, 2018{\natexlab{b}}.

\bibitem[Zhang et~al.(2019)Zhang, Khalili, and Liu]{zhang2019recycled}
Xueru Zhang, Mohammad~Mahdi Khalili, and Mingyan Liu.
\newblock Recycled admm: Improving the privacy and accuracy of distributed algorithms.
\newblock \emph{IEEE Transactions on Information Forensics and Security}, 15:\penalty0 1723--1734, 2019.

\bibitem[Zhang et~al.(2022)Zhang, Khalili, and Liu]{zhang2022differentially}
Xueru Zhang, Mohammad~Mahdi Khalili, and Mingyan Liu.
\newblock Differentially private real-time release of sequential data.
\newblock \emph{ACM Transactions on Privacy and Security}, 26\penalty0 (1):\penalty0 1--29, 2022.

\bibitem[Zhou et~al.(2018)Zhou, Moosavi-Dezfooli, Cheung, and Frossard]{dyn_quan_bit}
Yiren Zhou, Seyed-Mohsen Moosavi-Dezfooli, Ngai-Man Cheung, and Pascal Frossard.
\newblock Adaptive quantization for deep neural network.
\newblock In \emph{Proceedings of the AAAI Conference on Artificial Intelligence}, volume~32, 2018.

\end{thebibliography}

\newpage
\onecolumn

\title{Privacy-Aware Randomized Quantization via Linear Programming \\ (Supplementary Material)}
\maketitle

\appendix

\section{Proofs}\label{proofs}

\begin{proof}

(of Theorem~\ref{thm:erm_privacy}):

Assume that bins are uniformly distributed, i.e., $B_{i} = -\Delta-c + (i-1)\frac{2c+2\Delta}{m-1} ~ (i\in [m])$, $m \geq 4$, the selection probability of \textsf{ERM} can be calculated as:

\begin{align*}
    q_j(i) & = 
        \frac{\exp\{\frac{\gamma(B_{i}-B_{j})}{2(B_{j}-B_{1})}\}}{\sum_{k=1}^{j}\exp\{\frac{\gamma(B_{k}-B_{j})}{2(B_{j}-B_{1})}\}} 
         = \frac{\exp\{\frac{\gamma(i-j)}{2(j-1)}\}}{\sum_{k=1}^{j}\exp\{\frac{\gamma(k-j)}{2(j-1)}\}}.
\end{align*}

Take reciprocal of the probability, we have:

\begin{equation*}
    \frac{1}{q_j(i)} = \sum_{k=1}^{j} \exp\{\frac{\gamma(k-i)}{2(j-1)}\} 
      \leq j \exp \{ \frac{\gamma (j-1)}{2(j-1)} \} = j \exp\{\frac{\gamma}{2}\}. 
\end{equation*}

Therefore we can find the lower bound of $q_j(i)$:

\begin{equation*}
    q_j(i) \geq \frac{1}{j} \exp\{-\frac{\gamma}{2}\}.
\end{equation*}

Combining the lower bound with \eqref{equ:output}, we obtain a lower bound of $p(x,i)$:

\begin{align*}
    p(x,i) & = q_j(i)  {\sum}_{m\geq r \geq j+1} \left( q_{m-j}(m-r+1) \frac{B_{r}-x}{B_{r}-B_{i}}  \right) 
     ~~~\geq~~~ \frac{\exp\{-\frac{\gamma}{2}\}}{j} \cdot \frac{\exp\{-\frac{\gamma}{2}\}}{m-j} 
    \cdot {\sum}_{m\geq r \geq j+1} \left( \frac{B_{r}-x}{B_{r}-B_{i}}  \right) \\
    & \geq \frac{\exp\{-\gamma\}}{m^2/4} \cdot {\sum}_{m\geq r \geq j+1} \left( \frac{B_{r}-x}{B_{r}-B_{i}}  \right) 
     ~~~\geq~~~ \frac{4\exp\{-\gamma\}}{m^2 (B_{m}-B_{1})} \cdot {\sum}_{m\geq r \geq j+1} (B_{r}-B_{j+1}) \\
    & \geq \frac{4\exp\{-\gamma\}}{m^2 (2c+2\Delta)} \cdot \frac{2c+2\Delta}{m-1} \cdot {\sum}_{m\geq r \geq j+1} (r-(j+1)) 
     ~~~=~~~ \frac{4\exp\{-\gamma\}}{m^2 (m-1)} \cdot \frac{(m-j-1)(m-j)}{2}. 
\end{align*}

Since $x \in [-c, c]$ and $x \in [B_j, B_{j+1})$, we have:

\begin{equation*}
    (j-1) \cdot \frac{2c+2\Delta}{m-1} < \Delta \leq j \cdot \frac{2c+2\Delta}{m-1},
\end{equation*}

which implies:

\begin{equation*}
    \frac{(m-1)\Delta}{2c+2\Delta} \leq j < \frac{(m-1)\Delta}{2c+2\Delta} + 1.
\end{equation*}

Hence we have:

\begin{align*}
     p(x,i) & \geq \frac{4\exp\{-\gamma\}}{m^2 (m-1)} \cdot \frac{(m-j-1)(m-j)}{2} 
     ~~~>~~~  \frac{4\exp\{-\gamma\}}{m^2 (m-1)} \cdot \frac{(m-\frac{(m-1)\Delta}{2c+2\Delta}-2)(m-\frac{(m-1)\Delta}{2c+2\Delta} - 1)}{2} \\
    & =  \frac{(2mc+m\Delta-4c-4\Delta)(2mc+m\Delta-2c-2\Delta)\exp\{-\gamma\}}{2m^2 (m-1)(c+\Delta)}  
     ~~~>~~~  \frac{c\exp\{-\gamma\}}{2m(c+\Delta)} \quad (m \geq 4).
\end{align*}

Let privacy loss $e^\epsilon = \max_{x, x^{\prime}}\frac{p(x,i)}{p(x^{\prime},i)}$, $i \in [m]$, we have:

\begin{align*}
    e^\epsilon &
     \leq \frac{\max_{x,i} p(x,i)}{\min_{x,i} p(x,i)} \quad
     \leq \frac{1}{\frac{c\exp\{-\gamma\}}{2m(c+\Delta)}} 
= \frac{2m(c+\Delta)\exp{\gamma}}{c}. 
\end{align*}

Hence we have an upper bound on $\epsilon$:

\begin{equation*}
    \epsilon \leq \gamma + \log \frac{2m(c+\Delta)}{c}.
\end{equation*}

\end{proof}

\begin{proof}
(of Theorem~\ref{thm:erm_error}):

    For exponential mechanism with $\delta$-sensitive score function $f$, privacy parameter $\gamma$, set of output $\mathcal{Y}$, we have the following inequality on the quality $f(y)$ of the output $y$~\citep{adaptive}:

    \begin{align*}
        \mathbb{E}(f(y)) \geq {\max}_{y \in \mathcal{Y}} ~f(y) - \frac{2\delta\log |\mathcal{Y}|}{\gamma}.
    \end{align*}

    Assume $x \in [B_{j}, B_{j+1})$, $B_{i}=-c-\Delta+(i-1)\frac{2c+2\Delta}{m-1}$. When selecting the left bin $B_{l}$ with exponential mechanism (denote as event $L_j = l$), we have $f(l)=B_{j}-B_{l}$, $\max f(l) = 0$, sensitivity of the score function $\delta=B_{j}-B_{1}$, $|\mathcal{Y}|=j$, hence we have:

    \begin{align*}
        \mathbb{E}(B_{j}-B_{l}) = - \mathbb{E}(q(l)) \leq 
        \frac{2 (B_{j}-B_{1}) \log j}{\gamma}.
    \end{align*}

    Similarly, when selecting the right bin $B_{r}$, we have:

    \begin{align*}
        \mathbb{E}(B_{r}-B_{j+1}) \leq 
        \frac{(B_{m}-B_{j+1}) \log (m-j)}{\gamma}.
    \end{align*}

    Since $\mathcal{M}(x) \in \{B_{l}, B_{r}\}$, we can have an upper bound on the expected absolute error:

    \begin{align*}
         \mathbb{E}(|\mathcal{M}(x)-x|) & = \mathbb{E}(B_{r}-B_{j+1}) + (B_{j+1}-B_{j}) + \mathbb{E}(B_{j}-B_{l}) \\
        & \leq \frac{2 (B_{j}-B_{1}) \log j}{\gamma} + \frac{2c+2\Delta}{m-1} + \frac{(B_{m}-B_{j+1}) \log (m-j)}{\gamma} \\
        & \leq \frac{2 \log(m) (2c+2\Delta)}{\gamma} + \frac{2c+2\Delta}{m-1}.
    \end{align*}
    
\end{proof}

\begin{proof}
(of Lemma~\ref{lemma:x_error}):

For each given $x \in [B_j, B_{j+1})$, the upper bound of its Mean Absolute Error can be derived as follows:

{
\begin{eqnarray*}
    \mathbb{E}(|\mathcal{M}(x)-x|) &=&
    \sum_{\substack{i \leq j \\ k > j}} \Pr(L_j=i) \Pr(R_j=k) \big((\frac{B_{k}-x}{B_{k}-B_{i}}) (x-B_{i}) + (\frac{x-B_{i}}{B_{k}-B_{i}}) (B_{k}-x)\big)\\
     &=& \sum_{\substack{i \in [1, j] \\ k \in [j+1, m]}}\Pr(L_j=i)  \Pr(R_j=k) (\frac{2(x-B_{i})(B_{k}-x)}{B_{k}-B_{i}}) \\
     &\leq&  \sum_{\substack{i \in [1, j] \\ k \in [j+1, m]}}\Pr(L_j=i) \Pr(R_j=k) (\frac{B_{k}-B_{i}}{2}) \\
     &=&  \frac{1}{2}\mathbb{E}(B_{r}-B_{l}) \\ 
     &=& \frac{1}{2}\big(\mathbb{E}(B_{r}-B_{j+1}) + \mathbb{E}(B_{j+1}-B_{j}) + \mathbb{E}(B_{j}-B_{l})\big),
\end{eqnarray*}}
where $B_r$ is the random variable denoting the bin selected on the right, and $B_l$ is the random variable denoting the bin selected on the left.

Considering the process of selecting one bin from $n$ bins: $B_1, B_2, \cdots, B_n$ according to the selection probability $q_n(1), q_n(2), \cdots, q_n(n)$. Denote the expected distance between the output bin $B_{i}$ and $B_{n}$ as $\zeta_j$. $\zeta_n = \sum_{i \in [1,n]} q_n(i) (B_{n}-B_{i})$, which is the linear combination of selection probabilities when the value of bins are fixed. 
We know that $\mathbb{E}(B_{r}-B_{j+1}) = \zeta_{m-j}$, and $\mathbb{E}(B_{j}-B_{l}) = \zeta_{j}$.
Hence we obtain:
\begin{align*}
    \mathbb{E}(|\mathcal{M}(x)-x|) \leq \frac{1}{2} \big(\zeta_{m-j} + (B_{j+1}-B_{j}) + \zeta_{j}\big). \\
\end{align*}
\end{proof}

\begin{proof}
(of Theorem~\ref{thm:uniform}):

Assume that the position of bins are given (either uniformly or non-uniformly distributed), and the input $x \in [-c, c]$ follows uniform distribution, and the probability density function of $X$ is equal to $f_X(x) = \frac{1}{2c}$. Then, we find an upper bound  for $\mathbb{E}(|\mathcal{M}(X)-X|)$ using Lemma \ref{lemma:x_error} and law of total expectation as follows, 
{\small
\begin{align*}
     &\mathbb{E}(|\mathcal{M}(X)-X|) = \int_{-c}^c \frac{1}{2c}\mathbb{E}(|\mathcal{M}(x)-x|)dx \\  
     &= \int_{-c}^{B_s}\frac{1}{2c}\mathbb{E}(|\mathcal{M}(x)-x|)dx + \sum_{i=s}^{t-1}\int_{B_{i}}^{B_{i+1}}\frac{1}{2c}\mathbb{E}(|\mathcal{M}(x)-x|)dx +\int_{B_{t}}^{c}\frac{1}{2c}\mathbb{E}(|\mathcal{M}(x)-x|)dx \\
    &\leq  \frac{1}{2c} \big( (B_s+c)(\zeta_{m-s+1}+B_{s+1}-B_s+\zeta_{s-1})
    + \sum_{i=s}^{t-1} (B_{i+1}-B_{i})(\zeta_{m-i}+B_{i+1}-B_{i}+\zeta_{i}) 
    + (c-B_{t})(\zeta_{m-t}+B_{t+1}-B_{t}+\zeta_t) \big),
\end{align*}
}
 where $-c-\Delta \leq B_{s-1} < -c \leq B_s < B_{t} \leq c < B_{t+1}\leq c+\Delta $.
Discarding the constant terms, we can have the following objective function:
\begin{equation}
    \min_{q_j(i)} \sum_{s\leq n\leq  t+1} \big(\min(c, B_{n})-\max(-c, B_{n-1})\big) \big(\zeta_{n-1} + \zeta_{m-n+1}\big),
\end{equation}
where $\zeta_n$ is given in Lemma \ref{lemma:x_error} and Theorem~\ref{thm:uniform}.

\end{proof}

\begin{proof}
(of Lemma~\ref{lemma:pr_set}):

When $-c \leq B_i \leq c$ and $x \geq B_{i}$, according to \eqref{equ:output},  $p(x,i)=q_j(i)  {\sum}_{m\geq r \geq j+1} \left( q_{m-j}(m-r+1) \frac{B_{r}-x}{B_{r}-B_{i}}  \right)$. Since $\frac{B_{r}-x}{B_{r}-B_{i}} < 1$, we obtain that:

\begin{equation*}
    p(x,i) \leq q_j(i)  {\sum}_{m\geq r \geq j+1} q_{m-j}(m-r+1) = q_j(i). 
\end{equation*}

$\forall j \in [m], j \geq i$,  we assume $q_i(i) \geq q_{j}(i)$, hence we have $q_i(i) \in \overline{\mathcal{S}}_i$, where $\overline{\mathcal{S}}_i$ is as defined in Lemma~\ref{lemma:pr_set}. Similarly, we can prove that when $-c \leq B_i \leq c$ and $x < B_{i}$, $q_{m+1-i}(m+1-i) \in \overline{\mathcal{S}}_i$.

If $B_{i} \leq x$, then $\forall x, x^\prime \in [B_k, B_{k+1}), x \leq x^\prime$, we have $p(x,i) \geq p(x^\prime,i)$. This indicates that $p(x,i)$ is monotonic between each interval divided by bins (e.g., $[-c, B_{i})$, $[B_{i}, B_{i+1})$, or $[B_{i}, c)$), and is decreasing as $x$ is moving farther from $B_{i}$. We can also prove this when $x < B_i$. Hence if $B_i < -c$, $\max p(x,i)$ is achieved only when $x=-c$ or $x=B_{k}$ $(-c \leq B_{k} < c)$. If $B_{i} > c$, then $\max p(x,i)$ is achieved only when $x=c$ or $x=B_{k}$ $(-c \leq B_{k} < c)$. 
Similarly, $\min p(x,i)$ is achieved only when $x$ is approaching the position of bins $(B_{k})$, or locating at the edge ($c$ or $-c$) which is farther from $B_i$. 

\end{proof}

\begin{proof}
(of Theorem~\ref{thm:constraint2}):

According to Lemma~\ref{lemma:pr_set}, when $B_{i} \leq -c$, we have:

{ 
\begin{equation*}
\max_x p(x,i) \in \{p(-c,i)\} \cup \{p(B_{k},i)\} (k \in [m], -c \leq B_{k} \leq c).
\end{equation*}}

According to \eqref{equ:output}, when $B_{i} \leq x$:

{ 
\begin{equation*}
    p(x,i) = q_j(i)  {\sum}_{m\geq r \geq j+1} \left( q_{m-j}(m-r+1) \frac{B_{r}-x}{B_{r}-B_{i}}  \right) .
\end{equation*}}

Hence we can get:

{ 
\begin{equation*}
    p(B_{k},i) = q_k(i)  {\sum}_{m\geq r \geq k+1} \left( q_{m-k}(m-r+1) \frac{B_{r}-B_{k}}{B_{r}-B_{i}}  \right) .
\end{equation*}}

{ 
\begin{equation*}
    p(B_{k+1},i) = q_{k+1}(i)  {\sum}_{m\geq r \geq k+2} \left( q_{m-k-1}(m-r+1) \frac{B_{r}-B_{k+1}}{B_{r}-B_{i}}  \right) .
\end{equation*}}

Assume that $\forall i,j \in [m], i \leq j$:

{ 
\begin{equation}\label{equ:extra1}
    q_j(i) \geq q_{j+1}(i) , 
\end{equation}}

and $\forall k,r \in [m], s \leq k \leq t, r > k+1$ ($s$ and $t$ are as defined in Theorem~\ref{thm:uniform}), we assume:

\begin{equation*}
    q_{m-k}(m-r+1) \cdot (B_{r}-B_{k}) \geq q_{m-k-1}(m-r+1) \cdot (B_{r}-B_{k+1}),
\end{equation*}

then we get: 

{ 
\begin{equation}\label{equ:extra2}
    p(B_{k},i) \geq p(B_{k+1},i).
\end{equation}}

From \eqref{equ:output}, we can also know that:

{ 
\begin{equation*}
    p(x,i-1) = q_j(i-1)  {\sum}_{m\geq r \geq j+1} \left( q_{m-j}(m-r+1) \frac{B_{r}-x}{B_{r}-B_{i-1}}  \right) .
\end{equation*}}

Assume that $\forall i,j \in [m], i \leq j$, we have:
\begin{equation}\label{equ:extra3}
    q_j(i-1) \leq q_j(i),
\end{equation}

hence we can know that:

{ 
\begin{equation}\label{equ:extra4}
    p(x,i-1) \leq p(x, i). 
\end{equation}}

Through \eqref{equ:extra2}, \eqref{equ:extra4}, and Lemma~\ref{lemma:pr_set}, we can know that when $B_{i} < -c$, we have $p(-c, s-1) \in \overline{\mathcal{S}}_i$, where $\overline{\mathcal{S}}_i$ is as defined in Lemma~\ref{lemma:pr_set}, $s$ is as defined in Theorem~\ref{thm:uniform}.

When $-c \leq B_{i} \leq c$, we have $\max_x p(x,i) \in \{q_i(i), q_{m+1-i}(m+1-i)\}$. Assume that $\forall i \in [m], q_{i}(i) \geq q_{i+1}(i+1)$, we have $q_{s}(s) \in \overline{\mathcal{S}}_i$. Now we have $\overline{\mathcal{S}}_i = \{ p(-c, s-1), q_{s}(s) \}$.

According to Lemma~\ref{lemma:pr_set} and \eqref{equ:extra4}, $\min_{x,i} p(x,i) \in \{ \lim_{x \to B_{k}} p(x,1) |-c\leq B_{k} \leq c\} \cup \{p(c,1)\}$. We have:

{ 
\begin{equation}
    \lim_{x \to B_{k}}p(x,1)
    =  q_{k-1}(1)  {\sum}_{r \in [k+1, m]}\bigg( q_{m-k+1}(m-r+1) \frac{B_{r}-B_{k}}{B_{r}-B_{1}} \bigg) ,
\end{equation}}

{ 
\begin{equation}
    \lim_{x \to B_{k+1}}p(x,1)   
    =  q_{k}(1)  {\sum}_{r \in [k+2, m]}\bigg( q_{m-k}(m-r+1) \frac{B_{r}-B_{k+1}}{B_{r}-B_{1}} \bigg) .
\end{equation}}

According to \eqref{equ:extra1}, we have $q_{k-1}(j) \geq q_{k}(j)$, hence by requiring that for any $r, k  \in [m], s \leq k \leq t, r > k+1$:

{ 
\begin{equation}\label{equ:extra7}
    q_{m-k+1}(m-r+1)(B_{r}-B_{k}) \geq q_{m-k}(m-r+1)(B_{r}-B_{k+1}) ,
\end{equation}}

we obtain:

{ 
\begin{equation}\label{equ:extra7}
    \lim_{x \to B_{k}}p(x,1) >  \lim_{x \to B_{k+1}}p(x,1).
\end{equation}}

Combining \eqref{equ:extra7} with Lemma~\ref{lemma:pr_set}, we can know that:

\begin{align}
    \min_x p(x,i) \in \{ \lim_{x \to B_t} p(x,1), p(c,1)\}.
\end{align}

\end{proof}

\section{Experimental details}\label{exp_detail}

The hyperparameters used in each experiment are given as follows.

\begin{minipage}[t]{0.25\textwidth}

\begin{table}[H]\small
\setlength\tabcolsep{1pt}
    \centering
    \begin{tabular}{cc}
         \toprule
        Hyperparameter & Value \\
        \midrule
        OPTM bins & [-6.00, -0.40, 0.40, 6.00] \\
        MVU bins & [-4.34, -3.60, 3.60, 4.34] \\
        \bottomrule
    \end{tabular}
    \caption{Hyperparameters for scalar inputs when $\epsilon$ = 0.5}
    \label{tab:hp_eps0.5}
\end{table}
\end{minipage}
\qquad
\qquad
\begin{minipage}[t]{0.25\textwidth}
    \begin{table}[H]\small
    \setlength\tabcolsep{1pt}
    \centering
    \begin{tabular}{cc}
    \toprule
    Hyperparameter & Value \\
    \midrule
        OPTM bins & [-3.00, -0.50, 0.50, 3.00] \\
        MVU bins & [-2.42, -1.69, 1.69, 2.42] \\
        ERM $\gamma$ & 0.026 \\
        ERM bins & [-5.10, -0.10, 0.10, 5.10] \\
        RQM $q$ & 0.220 \\
        RQM bins & [-2.70, -0.90, 0.90, 2.70] \\
   \bottomrule
\end{tabular}
    \caption{Hyperparameters for scalar inputs when $\epsilon$ = 1.0}
    \label{tab:hp_eps1.0}
\end{table}
\end{minipage}
\qquad
\qquad
\begin{minipage}[t]{0.25\textwidth}
    \begin{table}[H]\small
    \setlength\tabcolsep{1pt}
    \centering
    \begin{tabular}{cc}
        \toprule
        Hyperparameter & Value \\
        \midrule
        OPTM bins & [-3.00, -0.50, 0.50, 3.00] \\
        MVU bins & [-1.83, -1.11, 1.11, 1.83] \\
        ERM $\gamma$ & 0.043 \\
        ERM bins & [-2.70, -0.40, 0.40, 2.70] \\
        RQM $q$ & 0.498 \\
        RQM bins & [-2.60, -0.87, 0.87, 2.60] \\
        \bottomrule
    \end{tabular}
    \caption{Hyperparameters for scalar inputs when $\epsilon$ = 1.5}
    \label{tab:hp_eps1.5}
\end{table}
\end{minipage}

\begin{minipage}[t]{0.25\textwidth}
    \begin{table}[H]\small
    \setlength\tabcolsep{1pt}
    \centering
    \begin{tabular}{cc}
    \toprule
    Hyperparameter & Value \\
    \midrule
        OPTM bins & [-4.00, 0.20, 0.60, 4.00] \\
        MVU bins & [-2.42, -1.69, 1.69, 2.42] \\
        RQM $q$ & 0.220 \\
        RQM bins & [-2.70, -0.90, 0.90, 2.70] \\
   \bottomrule
\end{tabular}
    \caption{Hyperparameters for truncated Gaussian distribution}
    \label{tab:hp_gauss}
\end{table}
\end{minipage}
\qquad
\qquad
\begin{minipage}[t]{0.25\textwidth}
\begin{table}[H]\small
\setlength\tabcolsep{1pt}
    \centering
    \begin{tabular}{cc}
         \toprule
        Hyperparameter & Value \\
        \midrule
        OPTM bins & [-3, -0.5, 0.5, 3] \\
        RQM bins & [-3, -1, 1, 3] \\
        \bottomrule
    \end{tabular}
    \caption{Hyperparameters for vector inputs}
    \label{tab:dp_vector}
\end{table}
\end{minipage}
\qquad
\begin{minipage}[t]{0.25\textwidth}
    \begin{table}[H]\small
    \setlength\tabcolsep{1pt}
    \centering
    \begin{tabular}{cc}
    \toprule
    Hyperparameter & Value \\
    \midrule
    Batch size & 8 \\
    DP budget $\epsilon$ & 1 \\
    Gradient norm clip & 0.1 \\
    OPTM bins & [-2.2, -0.4, 0.4, 2.2] \\
    RQM bins & [-2.7, -0.9, 0.9, 2.7] \\
    RQM $q$ & 0.22 \\
   \bottomrule
\end{tabular}
    \caption{Hyperparameters for DP-SGD on Breast Cancer dataset}
    \label{tab:dp_cancer}
\end{table}
\end{minipage}

\begin{minipage}[t]{0.25\textwidth}
    \begin{table}[H]\small
    \setlength\tabcolsep{1pt}
    \centering
    \begin{tabular}{cc}
        \toprule
        Hyperparameter & Value \\
        \midrule
        Batch size & 32 \\
        DP budget $\epsilon$ & 1 \\
        Gradient norm clip & 0.01 \\
        OPTM bins & [-2.6, -0.4, 0.4, 2.6] \\
        RQM bins & [-2.7, -0.9, 0.9, 2.7] \\
        RQM $q$ & 0.22 \\
        \bottomrule
    \end{tabular}
    \caption{Hyperparameters for DP-SGD on MNIST}
    \label{tab:dp_mnist}
\end{table}
\end{minipage}

\end{document}